\documentclass[sigconf]{acmart}
\copyrightyear{2026}
\acmYear{2026}
\setcopyright{cc}
\setcctype{by}
\acmConference[HPDC '26]{The 35th International Symposium on High-Performance Parallel and Distributed Computing}{July 13--16, 2026}{Cleveland, OH, USA}
\acmBooktitle{The 35th International Symposium on High-Performance Parallel and Distributed Computing (HPDC '26), July 13--16, 2026, Cleveland, OH, USA}
\acmDOI{10.1145/3806645.3807813}
\acmISBN{979-8-4007-2640-8/2026/07}

\usepackage{algorithm}
\usepackage[noend]{algorithmic}

\usepackage{subfigure}
\usepackage{multirow}
\usepackage{enumitem}
\usepackage{xspace}

\makeatletter
\newcommand{\removelatexerror}{\let\@latex@error\@gobble}
\makeatother

\newcommand{\eat}[1]{}

\newcommand{\stitle}[1]{\vspace{1ex} \noindent{{\bf #1}}}

\newcommand{\ours}[0]{STAR\xspace}

\begin{document}

\title[STAR: Decode-Phase Rescheduling for LLM Inference]{\ours: Decode-Phase Rescheduling for LLM Inference}         


\author{Zhibin Wang}
\affiliation{%
  \department{State Key Laboratory for Novel Software Technology,}
  \institution{Nanjing University}
  \city{Nanjing}
  \country{China}
}
\email{wzbwangzhibin@gmail.com}

\author{Zetao Hong}
\affiliation{%
  \department{State Key Laboratory for Novel Software Technology,}
  \institution{Nanjing University}
  \city{Nanjing}
  \country{China}
}
\email{mariohong128@gmail.com}

\author{Xue Li}
\affiliation{%
  \institution{Alibaba Group}
  \city{Hangzhou}
  \country{China}
}
\email{youli.lx@alibaba-inc.com}

\author{Zibo Wang}
\affiliation{%
  \department{State Key Laboratory for Novel Software Technology,}
  \institution{Nanjing University}
  \city{Nanjing}
  \country{China}
}
\email{wangzb@smail.nju.edu.cn}

\author{Shipeng Li}
\affiliation{%
  \department{State Key Laboratory for Novel Software Technology,}
  \institution{Nanjing University}
  \city{Nanjing}
  \country{China}
}
\email{spli@smail.nju.edu.cn}

\author{Qingkai Meng}
\affiliation{%
  \department{State Key Laboratory for Novel Software Technology,}
  \institution{Nanjing University}
  \city{Nanjing}
  \country{China}
}
\email{qkmeng@nju.edu.cn}

\author{Qing Wang}
\affiliation{%
  \department{State Key Laboratory for Novel Software Technology,}
  \institution{Nanjing University}
  \city{Nanjing}
  \country{China}
}
\email{wangqing.cs@nju.edu.cn}

\author{Chengying Huan}
\affiliation{%
  \department{State Key Laboratory for Novel Software Technology,}
  \institution{Nanjing University}
  \city{Nanjing}
  \country{China}
}
\email{huanchengying@nju.edu.cn}

\author{Rong Gu}
\authornote{Corresponding author.}
\affiliation{%
  \department{State Key Laboratory for Novel Software Technology,}
  \institution{Nanjing University}
  \city{Nanjing}
  \country{China}
}
\email{gurong@nju.edu.cn}

\author{Sheng Zhong}
\affiliation{%
  \department{State Key Laboratory for Novel Software Technology,}
  \institution{Nanjing University}
  \city{Nanjing}
  \country{China}
}
\email{sheng.zhong@gmail.com}

\author{Chen Tian}
\affiliation{%
  \department{State Key Laboratory for Novel Software Technology,}
  \institution{Nanjing University}
  \city{Nanjing}
  \country{China}
}
\email{tianchen@nju.edu.cn}

\renewcommand{\shortauthors}{Wang et al.}

\begin{abstract}
    Large Language Model (LLM) inference has emerged as a fundamental paradigm, however, variations in output length cause severe workload imbalance in the decode phase, particularly for long-output reasoning tasks. Existing systems, such as PD disaggregation architectures, rely on static prefill-to-decode scheduling, which often results in SLO violations and OOM failures under evolving decode workloads.
    In this paper, we propose \ours, a decode rescheduling system powered by length prediction to anticipate future workloads. Our core contributions include: (1) A lightweight and continuous LLM-native prediction method that leverages LLM hidden state to model remaining generation length with high precision (reducing MAE by 49.42\%) and low overhead (cutting predictor parameters by 93.28\%); (2) A rescheduling solution in decode phase with a dynamic balancing mechanism that integrates current and predicted workloads, reducing P99 TPOT by 75.1\% and achieving 2.63$\times$ higher goodput.
\end{abstract}



\maketitle

\section{Introduction}

With the great success of Large Language Models (LLMs) such as ChatGPT~\cite{openai2024chatgpt}, Claude~\cite{anthropic2024claude}, Qwen~\cite{qwenWeb}, Gemini~\cite{google2024gemini}, and DeepSeek~\cite{deepseekWeb}, the demand for LLM services is surging. Despite their success, the deployment of LLMs imposes heavy requirements on hardware resources, leading to high serving costs. Consequently, vast systems and algorithms~\cite{chen2024efficienteconomiclargelanguage,miao2024spotserve,gao2024cachedattention} are proposed to improve efficiency and reduce the cost of LLM serving.

\begin{figure}[!t]
    \centering
    \includegraphics[width=1.0\columnwidth]{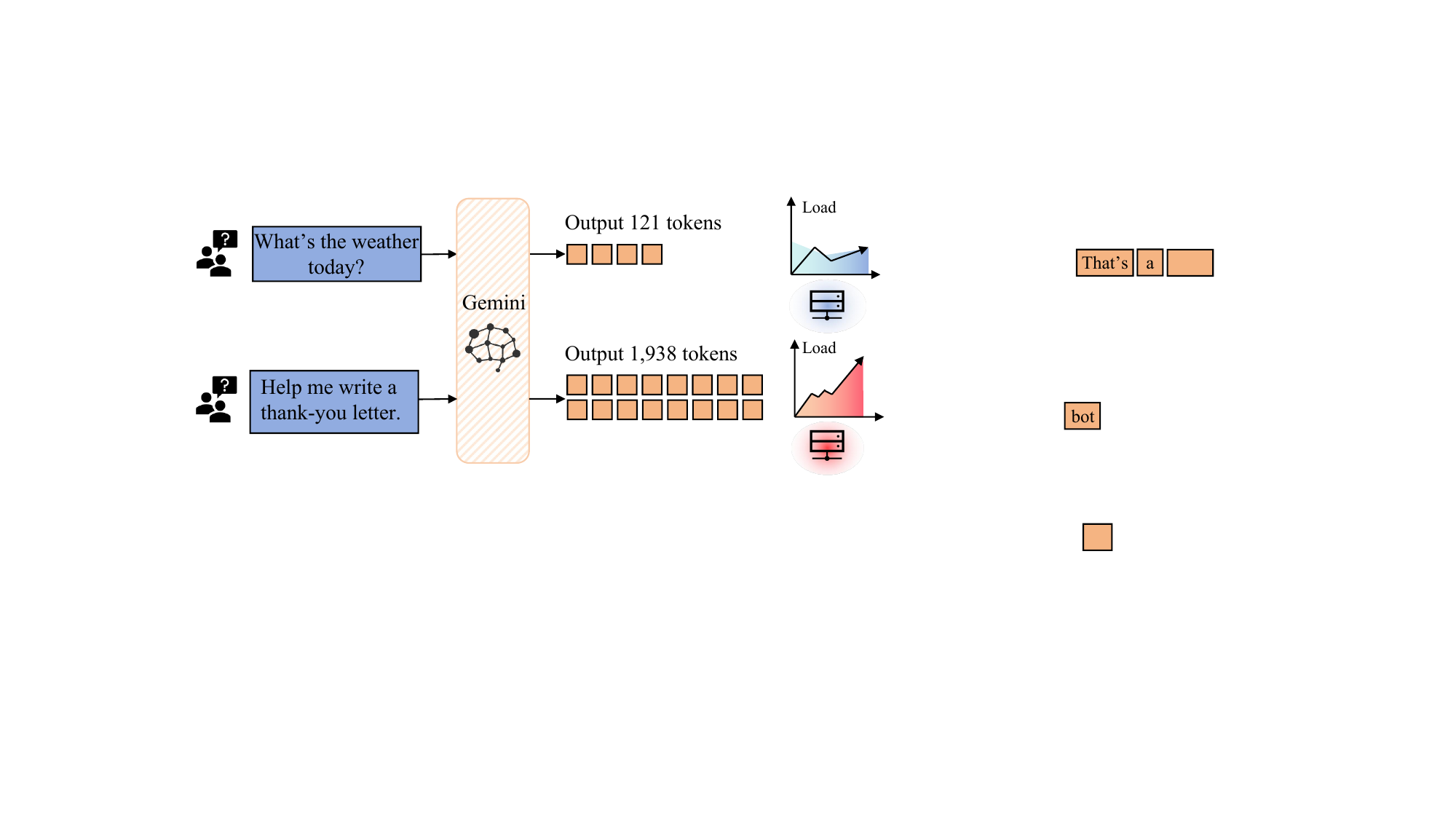}
    \caption{Different output lengths lead to significant load variations across decode instances.}
    \label{fig:first_example}
\end{figure}

The autoregressive inference process of LLMs results in significant workload variations, especially in reasoning tasks that require long chain-of-thought (CoT) outputs~\cite{wei2022chain}.
As illustrated in Figure~\ref{fig:first_example}, when using Gemini to process two different real-world inputs, the resulting output lengths differ by more than 16 times. The workload of LLM inference for a given model is determined by the length of the input and output, as inference of a request involves two stages: 1) input-related prefill processes the input prompt with a single forward pass to generate the initial output token and the key-value (KV) cache; 2) output-related decode generates output tokens auto-regressively, i.e., one token at a time, by leveraging the KV cache. As revealed in~\cite{deepseekai2024deepseekv3technicalreport}, the decode phase dominates the total cost in LLM inference, especially for long outputs. Therefore, the variation in output length leads to significant workload imbalance across requests during the decode phase.

Specifically, the variation of workload in the decode phase leads to two critical issues:
\begin{itemize}
    \item \textbf{Issue 1: OOM of KV cache.} In the decode phase, uneven distribution of workload across instances leads to overloaded instances accumulating large numbers of tokens, which causes their KV cache memory usage to rapidly grow. Although techniques like PagedAttention~\cite{kwon2023efficient} enable more flexible KV cache management, the total KV cache memory usage on overloaded instances can still exceed the hardware memory limit as long-output generations continue, resulting in ``OOM errors''. This risk also remains practical in real deployments, where concurrency is typically tuned for throughput rather than the worst-case KV-cache footprint of every request. For example, Anyscale explicitly warns that increasing the concurrent-sequence limit can cause OOM errors in long-context LLM serving~\cite{anyscale_llm_tuning}. Once OOM occurs, all affected requests on that instance are forced to recompute KV cache, greatly increasing their latency and reducing system throughput. Meanwhile, the overloaded (now OOM) instance's performance drops, its requests accumulate, and new requests are more likely to be assigned to other instances, causing further imbalance and overloading those as well. This chain reaction can produce cascading OOM errors and lead to overall system instability.
    \item \textbf{Issue 2: Violations of SLO.} Considering the distinct characteristics of prefill and decode, modern LLM serving systems~\cite{zhong2024distserve, mooncake} adopt a prefill-decode (PD) disaggregation architecture that separates the single, but relatively long, prefill iteration from the multiple, but shorter, decode iterations to prevent phase interference. Therefore, the time-per-output-token (TPOT) of decode will not be affected by the prefill phase. However, as the time to read the KV cache will increase linearly as the output length increases, long-output requests may lead to a significant increase in TPOT, which may violate the SLO of TPOT.
\end{itemize}

However, existing LLM serving systems~\cite{zhong2024distserve, mooncake, agrawal2023sarathiefficientllminference, yu2022orca, kwon2023efficient} are not prepared to handle workload imbalance posed by long output reasoning tasks. 1) Several works~\cite{mooncake} consider the load of each prefill as well as the cache of KV cache to \emph{assign the incoming request to a prefill instance}, but lack consideration of the long-term workload of decode instances. 2) Other works~\cite{vllmpd, li2025flowkvdisaggregatedinferenceframework} further consider the workload-balancing scheduling when \emph{dispatching requests from prefill to decode}: Round-robin scheduling~\cite{vllmpd} assigns requests to decode instances in a round-robin manner, ensuring an even distribution of requests, however, it overlooks the varying workloads of different requests; Current-load-balancing scheduling~\cite{li2025flowkvdisaggregatedinferenceframework} allocates requests based on the current load of each decode instance, i.e., size of KV cache, aiming to balance memory usage, but the evolving nature of requests results in imbalancing after a period of execution. Additionally, Llumnix~\cite{sun2024llumnix} proposes dynamic runtime rescheduling via live migration to improve isolation, mitigate fragmentation, and support priorities with strong tail-latency gains; however, it does not consider the prefill-decode disaggregation architecture, the decode characteristics under long outputs, nor future execution states (e.g., remaining length), relying mainly on current-state decisions.

In this paper, we propose to address the above limitations with the rescheduling in decode phase. Specifically, we aim to enable the migration of requests across decode instances during the decode phase to resolve workload imbalance, thereby improving SLO compliance and preventing OOM errors. However, implementing decode rescheduling presents two key challenges:

\stitle{Challenge 1: Accurate and efficient modeling of workload.}
In addition to the current workload of each decode instance, the scheduler must also consider the future decode workload contributed by currently active requests, which can be modeled by their remaining generation lengths.
However, accurately predicting the remaining generation length is challenging due to the high variance and unpredictability of output lengths in real-world scenarios. Existing methods either 1) rely on auxiliary models~\cite{qiu2024power, jin2023s, hu2024inference, fu2024efficientllmschedulinglearning}, which introduce additional computational overhead and limit by the ability of auxiliary models; or 2) inject tokens into the input prompt~\cite{zheng2023response} through prompt engineering, which is intrusive and may affect the quality of the generated output. Worse still, for same input prompt, the output will vary due to the temperature sampling and nondeterminism of LLM system, making accurate prediction with only input prompt impossible.

\stitle{Challenge 2: Effective rescheduling strategy in complex decision space.}
Unlike traditional task scheduling, decode rescheduling must consider both the current load of each instance and the remaining-token workload of active requests. It must also decide which request to migrate and when, since migrating a near-complete request may not amortize migration overhead. This enlarges the decision space relative to traditional prefill-to-decode scheduling.

In response to the above challenges, we introduce \ours, abbreviated for \underline{s}mart \underline{t}oken-length \underline{a}ware \underline{r}escheduling. As far as we know, \ours is the first system to consider decode rescheduling to resolve workload imbalance across decode instances. \ours achieves workload balancing through two key components:

\stitle{Lightweight and continuous LLM-native predictor (Section~\ref{sec:prediction})} efficiently and accurately estimates the remaining generation length of each request.
Specifically, instead of relying on auxiliary models or prompt engineering, we leverage the internal state of the original LLM to predict the remaining generation length. By feeding the hidden state of the last token from the final transformer layer into a lightweight MLP predictor, we achieve both lower prediction overhead and more accurate results.
Compared to the SOTA, our method reduces prediction mean absolute error (MAE) by 49.42\% on average while reducing overhead by 96.26\% for output lengths up to 32K tokens.
Moreover, we observe that by leveraging the additional generated tokens, we can further improve the prediction precision. With the low overhead of our LLM-native predictor, we can perform continuous prediction in an iterative manner, further enhancing precision in decision-making of decode rescheduling.

\stitle{Multi-stage rescheduling strategy (Section~\ref{sec:scheduler})} that effectively balances the workload across decode instances.
We first align the execution time and memory usage of each decode instance through the number of tokens in the batch, facilitating the further modeling of workload. Then, we design a multi-stage rescheduling strategy that 1) identifies overloaded and underloaded decode instances considering both current and predicted workload; 2) enumerates requests for each overloaded-underloaded instance pair and filters requests that can benefit from migration; 3) simulates the migration for each request and selects the best feasible migration that maximizes workload variance reduction. By periodically executing this rescheduling strategy, we can achieve dynamic workload balancing across decode instances, thereby improving SLO compliance and preventing OOM errors.

Comprehensive evaluations (Section~\ref{sec:evaluation}) demonstrate that \ours significantly outperforms SOTA LLM serving systems.
Compared to PD disaggregation implemented in vLLM~\cite{kwon2023efficient}, \ours achieves up to 2.63$\times$ higher goodput, reduces P99 TPOT latency by 75.1\%, and prevents OOM occurrences. Moreover, the simulation on larger-scale clusters shows that rescheduling effectively improves cluster load balancing, while prediction further reduces load fluctuations.

\section{Background}

\subsection{Prefill and Decode Phases}

Large language model inference has two phases: prefill and decode. Prefill processes the full input in one forward pass to generate the first token and build the KV cache; it is compute-bound and handled per request. Decode then generates tokens auto-regressively using the KV cache; it is memory-bound and typically batches requests for efficiency. Their SLOs differ: prefill minimizes Time-to-First-Token (TTFT), while decode reduces Time-per-Output-Token (TPOT).

Recognizing the distinct characteristics of these two phases, modern LLM serving systems (e.g., Mooncake~\cite{mooncake}, DistServe~\cite{zhong2024distserve}) separate prefill and decode onto different hardware resources to satisfy their respective resource demands.
Upon arrival, a request exclusively occupies a prefill instance, which queues inputs in FIFO order and selects instances based on load or KV cache reuse~\cite{mooncake}.
After the prefill phase, the request will be forwarded to a decode instance to generate tokens auto-regressively. Instead of queuing requests, decode instances batch multiple requests together as proposed in vLLM~\cite{kwon2023efficient}, Orca~\cite{yu2022orca} to improve hardware utilization.

Particularly, taking deployment of DeepSeek-V3~\cite{deepseekai2024deepseekv3technicalreport} as an example,
to fully utilize the hardware resources, a prefill pod leverages 32 H800 GPUs, while a decode pod utilizes 320 GPUs, which highlights that decode is the resource-consuming phase.
In the subsequent sections, we will delve deeper into the decode phase.

\subsection{Imbalance Workload in Decode Phase}
\label{subsec:workload_characteristics}

Given the autoregressive nature of decode, the workload of each request is determined by the output length. Therefore, we first analyze the characteristics of output length in real-world scenarios.

\begin{figure}[!b]
    \vspace{-0.06in}
    \centering
    \includegraphics[width=\linewidth,height=0.5\textwidth,keepaspectratio]{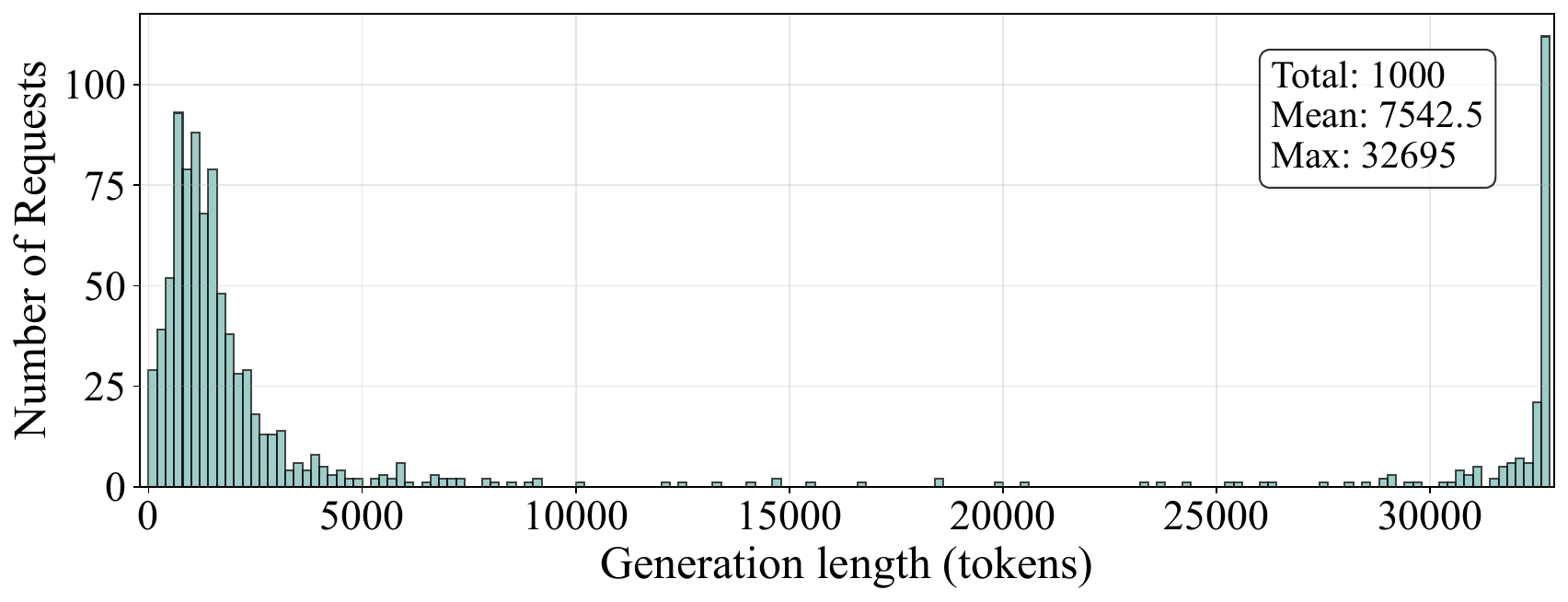}
    \caption{Output length distribution.}
    \label{fig: output_length_distribution}
\end{figure}

\begin{figure}[tbp]
    \vspace{-0.2in}
    \centering
    \subfigure[Round-robin scheduling]{
        \centering
        \includegraphics[width=.99\linewidth,height=0.6\textwidth,keepaspectratio]{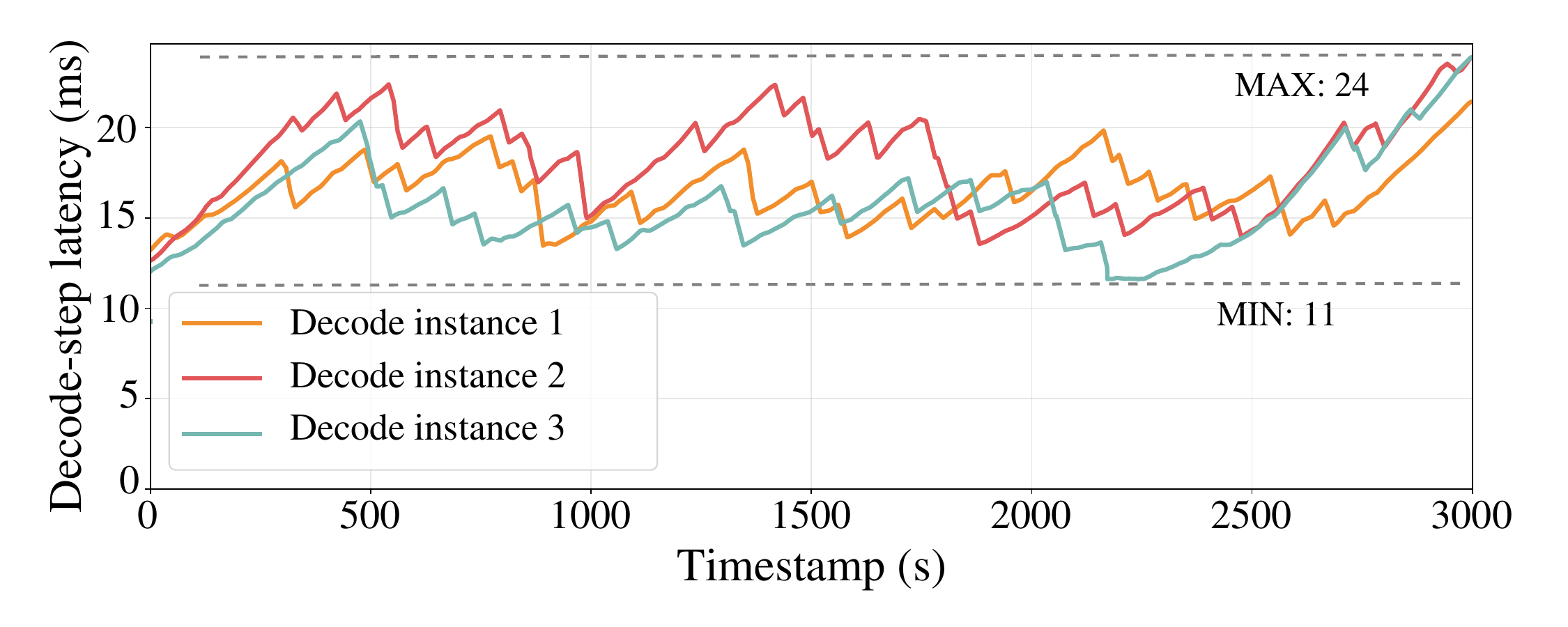}
        \label{figure_ca}}
    \subfigure[Current-load balancing]{
        \centering
        \includegraphics[width=.99\linewidth,height=0.6\textwidth,keepaspectratio]{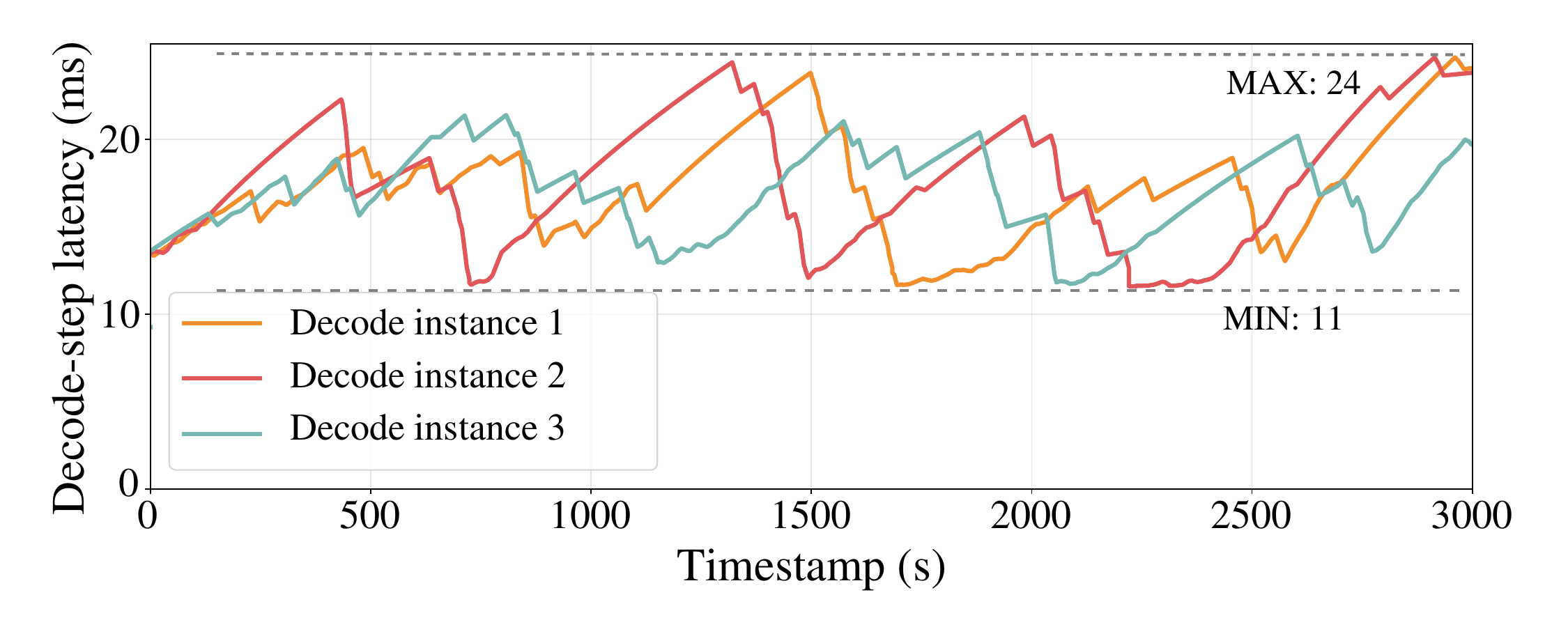}
        \label{figure_cb}
    }
    \caption{Per-instance decode-step latency over time across three decode instances under PD disaggregation (1 prefill + 3 decode instances).}
    \label{fig:tpot_variation}
\end{figure}

\stitle{Output length variation.} Figure~\ref{fig: output_length_distribution} illustrates the output length distribution in ShareGPT dataset when running DeepSeek-R1-Distill-Qwen-7B model with a maximum output length of 32K tokens. Many requests present short outputs, with 29.2\% requests generating fewer than 1K tokens, while a non-ignorable portion of requests produce very long outputs (17.3\% with exceeding 30K tokens).

\stitle{Existing prefill-to-decode scheduling.}
While PD disaggregation effectively eliminates phase interference, it introduces a critical limitation: it relies on statically assigning decode instances to achieve load balance. However, relying solely on prefill-to-decode scheduling is insufficient to address the workload imbalance within the decode phase, as it does not adapt to the varying computational and memory demands across decode instances.

\begin{itemize}
    \item \textbf{Round-robin scheduling~\cite{vllmpd}:} This straightforward approach assigns requests to decode instances in a round-robin manner, ensuring an even distribution of requests. However, it overlooks the varying workloads of different requests, leading to potential load imbalances.
    \item \textbf{Current-load balancing~\cite{li2025flowkvdisaggregatedinferenceframework}:} This method allocates requests based on the current KV cache size of each decode instance, aiming to balance memory usage. While it considers the memory footprint, it still fails to account for the actual computational load associated with generating output tokens, which can vary significantly between requests.
\end{itemize}
Figure~\ref{fig:tpot_variation} illustrates the TPOT variation across three decode instances under different prefill-to-decode scheduling strategies with the same workload in Figure~\ref{fig: output_length_distribution} and request rate of 0.1 per second. Even with initial load balancing during prefill-to-decode handoff, significant TPOT divergence emerges with the progression of generation. Decode instances exhibit rapidly escalating performance disparities as output sequences lengthen—requests with prolonged residency on a single instance dominate its resources, causing cascading TPOT spikes for subsequent requests.

\stitle{Summarization and Motivation.}
Existing prefill-to-decode scheduling strategies cannot effectively balance the workload across decode instances, especially in scenarios with long and variable output lengths. Rescheduling during the decode phase is necessary to dynamically adjust to workload variations.

\subsection{Generation Length Prediction}
\label{subsec:prediction_analysis}
\begin{figure}[t]
    \vspace{-0.12in}
    \centering
    \includegraphics[width=.9\columnwidth]{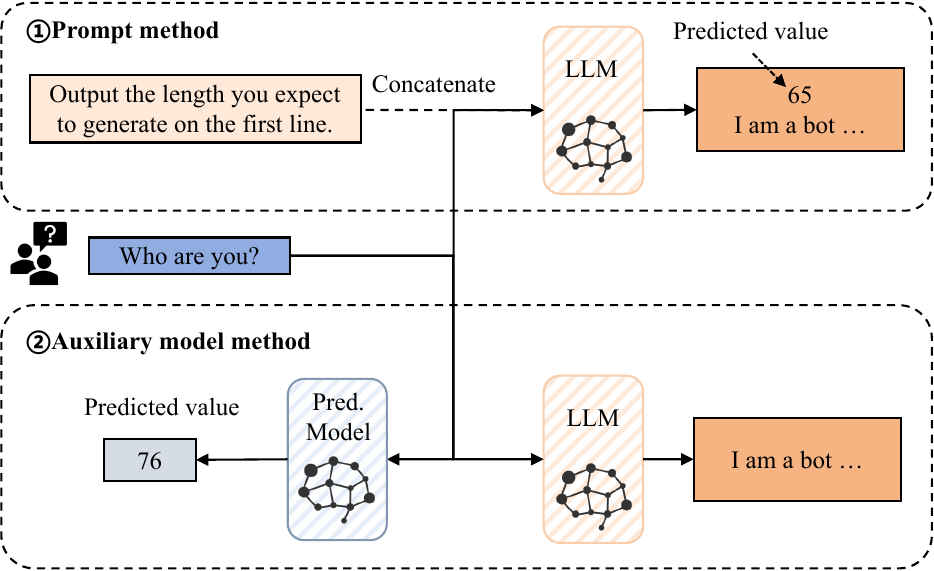}
    \caption{Current prediction methods.}
    \label{fig:current_prediction_methods}
\end{figure}

Accurate prediction of remaining output length is critical for modeling future workload in LLM inference systems. There are two main approaches—prompt method and auxiliary model method—as well as an additional iterative refinement method.

\begin{itemize}
    \item \textbf{Prompt-based methods} such as Perception in Advance (PiA) ~\cite{zheng2023response} modify user instructions to have the LLM first predict its output length before generating the response (Figure~\ref{fig:current_prediction_methods}). While achieving reasonable prediction accuracy, this approach requires intrusive modifications to user prompts, potentially altering model behavior and output quality.\cite{fu2024efficientllmschedulinglearning} As LLM applications have matured, such user-facing interventions have become increasingly unacceptable, rendering this approach impractical for modern serving systems.

    \item \textbf{Auxiliary model methods} employing smaller models like opt or bert (Figure~\ref{fig:current_prediction_methods}) with added prediction heads~\cite{jin2023s,qiu2024power,qiu2024efficient}. Sequence Scheduling~\cite{zheng2023response} established that prediction accuracy correlates with model capability, yet these auxiliary models are orders of magnitude smaller than target LLMs, fundamentally limiting their contextual understanding and prediction accuracy. Our evaluation shows that MAE increases by 3.2$\times$ when context length grows from 4K to 32K tokens, making these methods particularly ineffective for long output sequences.
\end{itemize}

\stitle{Summarization and Motivation.} Existing methods still suffer from poor accuracy and high overhead. This is arising from lack of consideration of integrating the prediction model with the target LLM. How to leverage the target LLM's own capabilities to provide accurate predictions with minimal overhead remains unsolved.

\subsection{Challenges}
In summary, to balance the inference workload across decode instances in PD disaggregation, we identify two key challenges:

\begin{itemize}
    \item \textbf{Accurate and efficient prediction of remaining output length:} Existing methods either require intrusive prompt modifications, rely on less capable auxiliary models with poor accuracy for long outputs, or incur prohibitive overheads that prevent iteration-level application. A new approach is needed that leverages the target LLM's own capabilities to provide precise predictions with minimal computational cost.
    \item \textbf{Complex dynamic rescheduling in decode phase:} Current PD disaggregation systems lack mechanisms for runtime migration between decode instances once requests are initially assigned. Extending traditional prefill-to-decode scheduling to enable such dynamic rescheduling is non-trivial, since decode-to-decode migration must: 1) account for the future system state, i.e., the predicted remaining output lengths of active requests, and 2) operate over a more complex decision space—not only deciding which decode instance to migrate to, as in prior work, but also when to migrate and which request to evict from a decode instance.
\end{itemize}

\section{Overview}
\begin{figure}[t]
    \centering
    \vspace{-0.12in}
    \includegraphics[width=1\columnwidth]{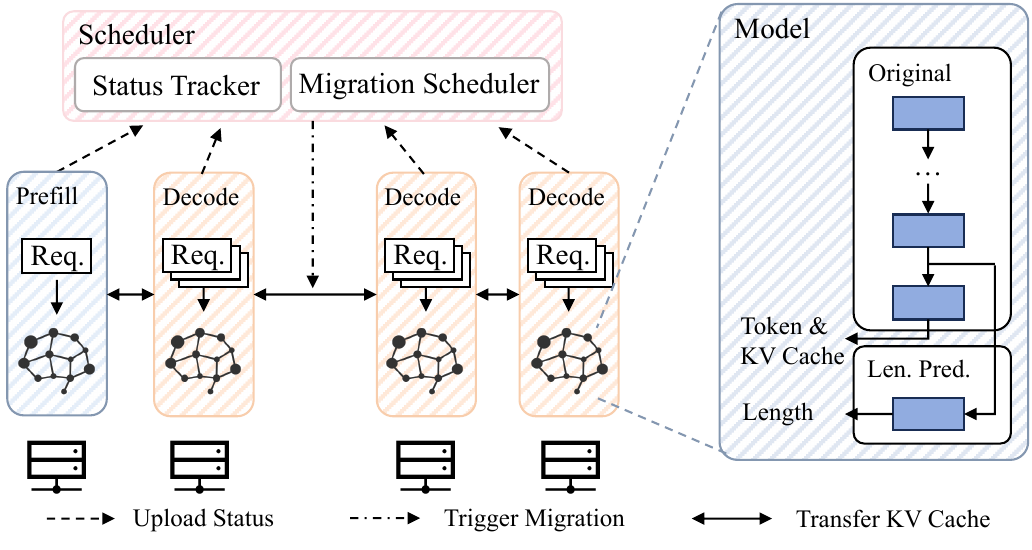}
    \caption{System overview.}
    \label{fig:overview}
\end{figure}
Figure~\ref{fig:overview} illustrates the overview of \ours built on the PD disaggregation architecture. When a request arrives, it is first sent to a prefill instance to process the input prompt and generate KV cache. After the prefill phase is completed, the request will be forwarded to a decode instance according to its input length, predicted output length, and the current load of each decode instance. During the decode phase, each request will be continuously predicted for its remaining output length at regular intervals, and the status of each decode instance will be reported to the scheduler. The scheduler will run the rescheduling algorithm to make migration decisions and trigger the migration of requests between decode instances. Particularly, \ours incorporates two novel components:

\stitle{Length predictor (Section~\ref{sec:prediction})} is used to forecast the model's output length, providing a reference for future status in scheduling algorithms. Instead of using the suboptimal approaches discussed in Section~\ref{subsec:prediction_analysis}, we leverage the model's internal state by feeding the hidden state of the last token from the final transformer layer into a lightweight MLP predictor.
Moreover, we continuously predict the remaining output length of each request at regular intervals during the decode phase to further improve prediction accuracy.

\stitle{Decode rescheduler (Section~\ref{sec:scheduler})} not only plays a role in task distribution but also handles the rescheduling of decode tasks.
To achieve this functionality, we introduce a multi-stage rescheduling approach. The process begins by evaluating both current and predicted workloads received from instances to identify instances that are either overloaded or underloaded. Next, we assess each instance pair and filter out requests that cannot benefit from migration. Finally, migration scenarios are simulated, and the best feasible migration is selected based on its ability to reduce workload variability. This rescheduling strategy is executed periodically, facilitating improved SLO adherence and minimizing the risk of OOM errors, ultimately enhancing system efficiency.

\section{Generation Length Prediction}
\label{sec:prediction}
The generation length predictor is used to provide a reference for future status in decode rescheduling decisions to select requests for migration. Our analysis in Section\ref{subsec:prediction_analysis} reveals existing prediction methods suffer from a fundamental accuracy ceiling because they rely on auxiliary models (e.g., bert-base-uncased, opt-125m) significantly less capable than target LLMs. As Sequence Scheduling~\cite{zheng2023response} established, prediction accuracy correlates with model capability.

\subsection{Overview of Predictor}
Contrary to prior works that perform generation length prediction only once after prefill, the prediction task in decode rescheduling presents following unique characteristics.

\stitle{Tight overhead constraints}:
In contrast to the loose TTFT constraints in prefill (e.g., 4s for Chatbot OPT-175B in DistServe~\cite{zhong2024distserve}), decode phase has stringent TPOT requirements (e.g., 0.2s for the same model).
Heavy auxiliary models or prompt modifications that reprocess the full context are therefore impractical.
As sequence length $l$ grows, such methods incur an extra $O(l)$ cost compared to the target LLM.
Unfortunately, as shown in Section~\ref{subsec:workload_characteristics}, longer sequences are common in real-world scenarios and is the key reason for developing \ours.

\stitle{Additional context from generated tokens}:
Rescheduling occurs during decode, after some output tokens have already been generated.
These tokens provide additional context that can be leveraged to improve prediction accuracy.
Existing methods are mismatched to this setting because they predict only once before decode and use only the prompt.

\begin{figure}[t]
    \centering
    \includegraphics[width=1.0\columnwidth]{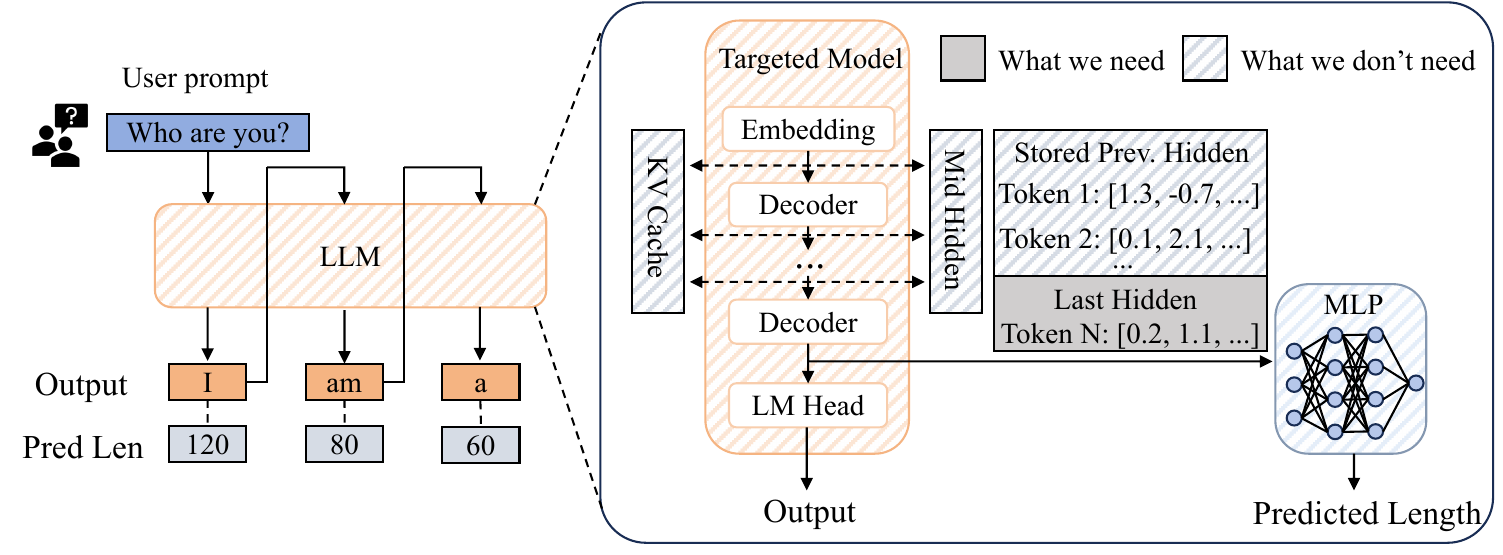}
    \caption{Our runtime prediction method: the MLP predictor consumes the hidden state vector of the last token from the final layer to estimate remaining output length. }
    \label{fig:prediction_method}
\end{figure}

Figure~\ref{fig:prediction_method} illustrates our proposed prediction method, which considers the above characteristics throughout two key design aspects: 1) leveraging the target LLM's own internal state, i.e., the hidden state of the last token from the final transformer layer, to achieve both low overhead and high accuracy (Section~\ref{subsec:lightweight_prediction_method}); and 2) continuously refining predictions according to the newly generated tokens during decode, thereby enhancing accuracy over time (Section~\ref{subsec:iterative_refinement}).

\subsection{Lightweight Target-LLM-Native Prediction}
\label{subsec:lightweight_prediction_method}

\stitle{Prediction Formulation.}
The prediction task can be formulated as a regression task.
Let $\vartheta$ be the parameters of the target LLM model, given an input sequence $S$ including prompt and generated tokens, we aim to learn a prediction function $f_\theta$ to estimate the remaining output length,
\begin{equation}
    f_\theta : (S) \mapsto p_\theta(Y \mid S),
\end{equation}
where $Y$ is the remaining output length and $\theta$ represents the model parameters of length predictor.

We observe that
\begin{center}
    \framebox{
        \begin{minipage}{0.9\linewidth}
            \emph{Length predictor $\theta$ can leverage the partial knowledge of target LLM $\vartheta$.}
        \end{minipage}
    }
\end{center}

Instead of using an auxiliary model, we propose a self-aware prediction method that leverages the target LLM's own hidden states to predict remaining output length.

\stitle{Design Choices.}
However, there are several design choices to consider when selecting the internal state as input to the prediction model, we discuss the design space in detail below.

\begin{itemize}
    \item \textit{KV cache vs. Hidden states:} Both KV cache and hidden states are part of the model's internal state. While their size and structure are similar, the information they carry differs: a token's KV cache contains only the information specific to that token, while a layer's hidden state encapsulates information from both the current token and other tokens. As a result, we believe the hidden state carries more informative content than the KV cache. Therefore, we choose the hidden state as the input for our model, a choice also seen in EAGLE-style methods, although those works use hidden states for speculative decoding rather than remaining-length prediction~\cite{li2025eaglespeculativesamplingrequires, li2024eagle2fasterinferencelanguage, li2025eagle3scalinginferenceacceleration}.
    \item \textit{All layers vs. last layer:} While we could choose to use the hidden state from all layers or just the last layer as the model input, aggregating hidden states from all layers increases both input dimensionality and computational cost. Since the last layer's hidden state already integrates information from all previous layers and captures the most abstract semantic features, using only the last layer strikes a good balance between information richness and efficiency.
    \item \textit{All tokens vs. last token:} Using all token states would provide more information, but it makes the input grow with sequence length and would require storing or recomputing preceding states. In contrast, the last token's state is already available at each decode step and, after attending to the full sequence, still captures the overall context. We therefore use only the last token's state as an efficient approximation.

\end{itemize}

\stitle{Our Solution: LLM-native Predictor.}
As demonstrated in Figure~\ref{fig:prediction_method}, our solution employs the hidden state of the last token from the last transformer layer as input to a lightweight MLP predictor.
This design ensures a fixed-size input that captures the model's full contextual understanding without incurring the high cost of storing every token's hidden state or processing larger states like the full KV cache.

\stitle{MLP Predictor.}
Formally, the predictor receives the last-layer hidden state of the last generated token as a fixed-size vector $h\in\mathbb{R}^{d}$ and outputs a scalar representing the remaining length through a 4-layer MLP:
\begin{equation}
    \hat{y} = w_4 \phi(\mathbf{W}_3 \phi(\mathbf{W}_2 \phi(\mathbf{W}_1 h))),
\end{equation}
where $\mathbf{W}_1 \in \mathbb{R}^{m_1 \times d}$, $\mathbf{W}_2 \in \mathbb{R}^{m_2 \times m_1}$, $\mathbf{W}_3 \in \mathbb{R}^{m_3 \times m_2}$, $w_4 \in \mathbb{R}^{1 \times m_3}$ are learnable parameters, and $\phi$ is an element-wise nonlinearity (ReLU). In the evaluated DeepSeek-R1-Distill-Qwen-7B model with $d=3584$, we set the intermediate dimensions to $m_1=2048$, $m_2=512$, and $m_3=64$.

\subsection{Enhancing Precision with Continuous Prediction}
\label{subsec:iterative_refinement}

Though our LLM-native prediction method achieves high accuracy compared to prior approaches, the absolute MAE of 3,873 tokens remains significant for long-output requests.
It is worth noting that prediction based solely on the prompt cannot achieve high accuracy, as the same input prompt can lead to highly variable output lengths due to the inherent randomness in LLM generation, including sampling strategies~\cite{holtzman2019curious}, temperature settings~\cite{li2025exploringimpacttemperaturelarge}, and system non-determinism~\cite{he2025nondeterminism}.
To further enhance prediction precision, we propose to continuously predict the remaining output length with the integration of additional context from generated tokens. This design is arised from the following observation:

\begin{center}
    \framebox{
        \begin{minipage}{0.9\linewidth}
            \emph{Additional context from generated tokens can be employed to improve prediction accuracy.}
        \end{minipage}
    }
\end{center}

These additional contexts are particularly valuable for two reasons: 1) they provide direct evidence of the model's current trajectory, allowing the predictor to adjust its estimate based on the actual content being generated; 2) richer context enables the predictor to capture nuanced patterns and dependencies that may not be evident from the prompt alone, leading to more informed and accurate predictions.

\begin{figure}[t]
    \vspace{-0.12in}
    \centering
    \includegraphics[width=1\columnwidth]{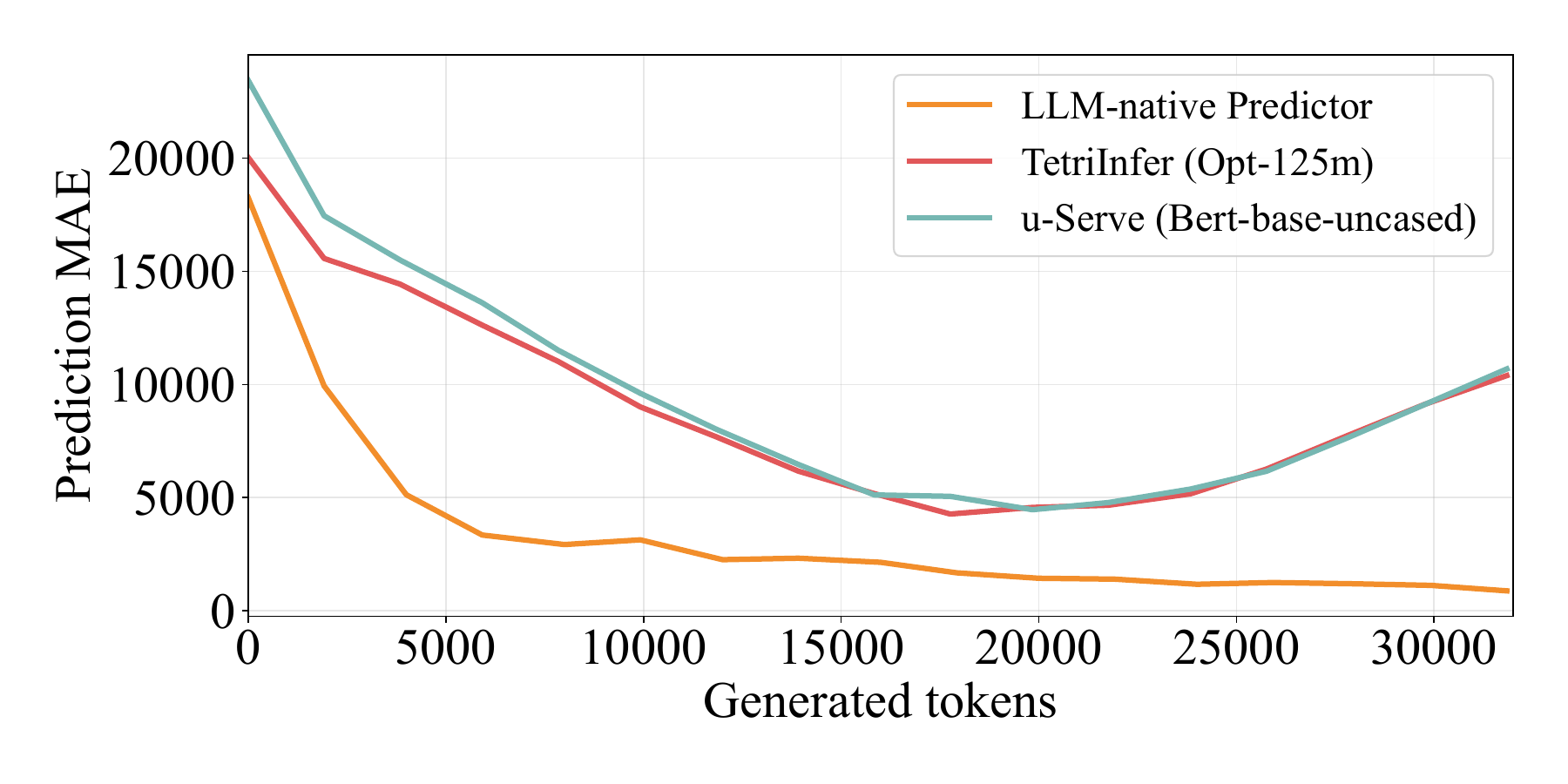}
    \caption{MAE of each prediction model for requests with 30-32K output tokens at different generated tokens.}
    \label{fig:prediction}
\end{figure}

\stitle{Empirical Evidence.}
Figure~\ref{fig:prediction} shows the prediction errors for requests that actually generate 30-32K tokens, measured at different generation stages (i.e., when different numbers of tokens have been generated).
The detailed experimental setup is described in Section~\ref{subsec:prediction_evaluation}.
With the increase of generated tokens, the amount of information fed into the prediction model increases, leading to precision of the prediction increasing. Specifically, with only input prompt, the MAE of our method is 18256 tokens, while after generating 8000 tokens, the MAE drops to 2929 tokens.

We notice that the existing auxiliary model methods, i.e., opt and bert, present an increasing trend of MAE as the number of generated tokens increases from 20000 to 30000. This is because these methods only support limited input lengths (1024 tokens for opt and 512 tokens for bert), thereby truncating long inputs, which results in a loss of information and a significant decrease in precision.

\subsection{Implementation and Empirical Evaluation}
\label{subsec:prediction_evaluation}
\stitle{Dataset Construction.} We construct a supervised dataset by running the vLLM engine on ShareGPT requests with the DeepSeek-R1-Distill-Qwen-7B model. For each request, we record the last-layer, last-token hidden state $\mathbf{h}_t$ together with the ground-truth remaining length $y_t$ at fixed decode intervals (e.g., every 20 tokens).
This yields 100k samples $\mathcal{D}=\{(\mathbf{h}_t, y_t)\}$ across the generation trajectory of each request.
To ensure proper evaluation, we split the data at the request level rather than the sample level.
Specifically, we randomly partition the original ShareGPT requests into training (70\%), validation (15\%), and test (15\%) sets.
This ensures that samples from the same request (at different generation timesteps) never appear in different splits, preventing data leakage and ensuring that the model is evaluated on truly unseen requests.

\stitle{Training Details.}
Both our LLM-native predictor and auxiliary models, i.e., $\mu$-Serve (bert-base-uncased)~\cite{qiu2024power} and TetriInfer (opt-125m)~\cite{hu2024inference}, are fine-tuned on the same training data on a single H800 GPU for up to 100 epochs (with early stopping patience of 10), minimizing a robust regression L1 loss using the AdamW optimizer~\cite{loshchilov2019decoupledweightdecayregularization} and employing early stopping~\cite{prechelt2002early} based on validation MAE.
Regarding MLP in our LLM-native predictor, we perform hyperparameter search over network depth (2-6 layers), hidden dimensions (256-1024), learning rate (1e-4 to 1e-3), and batch size (16-128), selecting the configuration with the best validation MAE.

\stitle{Result.} Table~\ref{tab:mae_overhead} summarizes the prediction accuracy and overhead of different methods, including 1) PiA~\cite{zheng2023response}, a prompt-based method that modifies user instructions to have the LLM first predict its output length before generating the response; 2) $\mu$-Serve~\cite{qiu2024power}, which uses bert-base-uncased as an auxiliary model; 3) TetriInfer~\cite{hu2024inference}, which uses opt-125m as an auxiliary model and 4) our LLM-native predictor. Regarding the training time, our LLM-native predictor presents a significant advantage compared to auxiliary models, while prompt-based methods PiA is training-free. Therefore, though prompt-based methods have large model size, lack of fine-tuning makes them poor in precision compared to trained models, while our LLM-native predictor achieves the best MAE. Regarding the inference latency, our LLM-native predictor is significantly smaller than auxiliary models and original LLM used in prompt-based methods, leading to the lowest latency.

\begin{table}[tbp]
    \centering
    \caption{Comparison of various generation length prediction methods.}
    \label{tab:mae_overhead}
    \small
    \begin{tabular}{l|cccc}
        \toprule
        \textbf{Methods}   & \textbf{PiA} & \textbf{$\mu$-Serve} & \textbf{TetriInfer} & \textbf{LLM-native} \\
        \midrule
        Parameters         & 7\,B         & 110\,M               & 125\,M              & 8.4\,M              \\
        Training time      & 0            & \~1\,day             & \~1\,day            & \~1\,hour           \\
        \midrule
        Average MAE        & 14169.3      & 8165.8               & 7658.1              & 3873.2              \\
        Latency (batch:1)  & 2.2\,s       & 6.0\,ms              & 10.3\,ms            & 1.33\,ms            \\
        Latency (batch:10) & 15.3\,s      & 30.0\,ms             & 65.3\,ms            & 2.4\,ms             \\
        \bottomrule
    \end{tabular}
\end{table}

\section{Decode Rescheduling}
\label{sec:scheduler}
With the help of LLM-native predictor described in Section~\ref{sec:prediction}, we can model the future state of the system and proactively schedule migrations between decode instances. This enables us to redistribute requests in a way that balances resource usage and minimizes latency spikes, especially under heavy or skewed workloads.

\subsection{From Prefill-to-Decode Scheduling to Decode-to-Decode Rescheduling}
In traditional prefill to decode scheduling, once a request finishes prefill, the scheduler assigns it to a decode instance based on system state and request characteristics at that moment. However, the rescheduling in decode phase presents distinct decision-making space as follows.

\stitle{Whether to migrate:}
Decode rescheduling makes migration optional. The scheduler can skip migration when the system is already balanced, when the benefit is smaller than the migration overhead, or when the overload is likely to disappear soon.

\stitle{Which request to migrate out:}
Decode rescheduling must explicitly choose a request to move out of an overloaded instance. This is a tradeoff: long requests relieve more load but incur larger KV-cache transfer and may overload the target, while short requests are cheaper to move but may not reduce enough load, especially if they are close to completion.

\stitle{Which instance to migrate in:}
The scheduler must also choose the destination instance for the migrated request. This decision depends on the load and available resources of candidate decode instances, together with the demand of the migrated request.

However, recognizing the evolving workload imbalance during decode phase, the scheduler needs to further consider the \emph{future state} of the system when making migration decisions.
Here, the future state is derived from the predicted remaining lengths of currently active decode requests only, rather than from future arrivals or future requests for the same query.

\subsection{Migration Algorithm}

To address workload imbalance described in Section~\ref{subsec:workload_characteristics} to ensure SLO compliance in decode instances, we design a migration algorithm that leverages execution time modeling and runtime prediction to make informed, low-overhead migration decisions. Before delving into the algorithm, we first review the workload characteristics in decode instances.

\begin{figure}[t]
    \vspace{-0.12in}
    \centering
    \subfigure[Decode execution time.]{
        \includegraphics[width=0.24\textwidth]{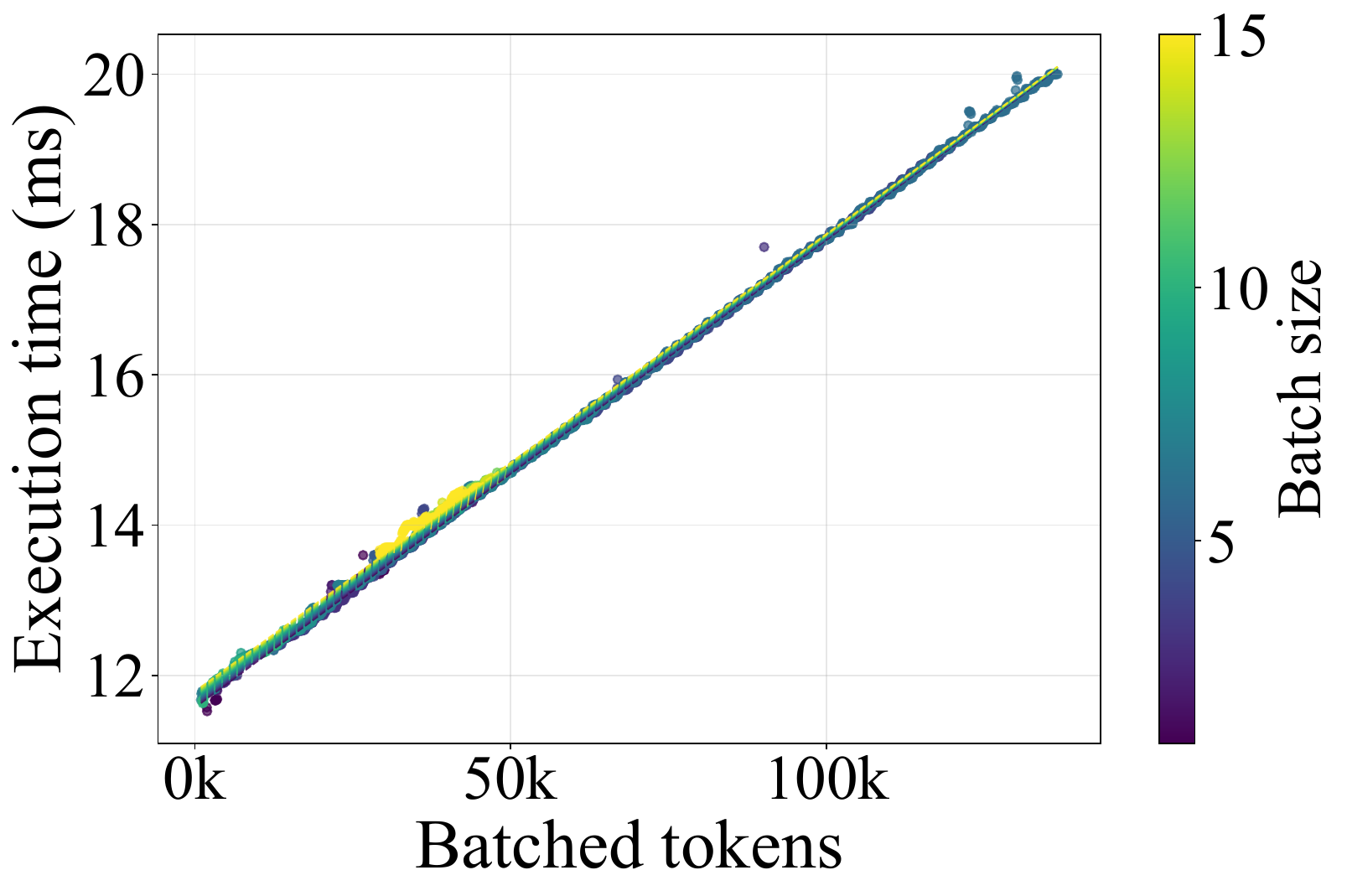}
    }
    \hfill
    \subfigure[KV cache usage.]{
        \includegraphics[width=0.2\textwidth]{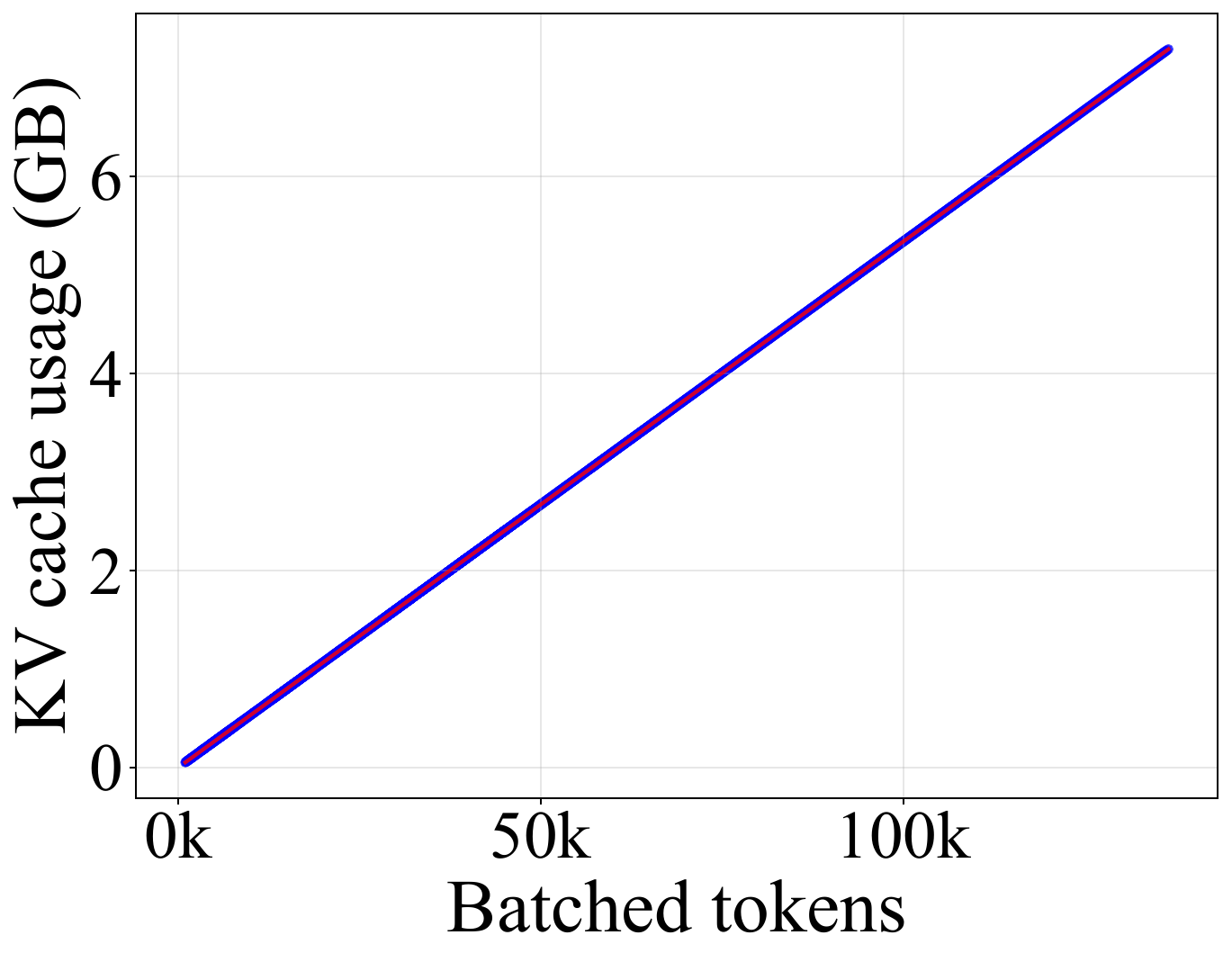}
    }
    \caption{Relationship between cost metrics and number of batched tokens.}
    \label{fig:performance_analysis}
\end{figure}

\stitle{Aligning Execution Time and Token load.}
Recently, researchers have conducted extensive studies on modeling the decode execution time of LLMs~\cite{cheng2025slicelevelschedulinghighthroughput}, either through analytical models or empirical benchmarking.
Despite the complexity of these models, one critical impact factor is the number of tokens in the running batch, which directly affects the time to read the KV cache during attention computation.
As shown in Figure~\ref{fig:performance_analysis}, the decode execution time per iteration is linearly correlated with the number of batched tokens (left panel).
The key reason is that the attention computation time is dominated by the KV cache read time~\cite{yu2022orca}, which grows linearly with the number of tokens in the batch.
Interestingly, another important factor, memory usage, is also linearly correlated with the number of tokens in the batch, as each token contributes a fixed-size KV cache (right panel).
Therefore, in this paper, we employ the \emph{number of tokens} in the running batch to unify the modeling of both execution time and memory usage, i.e., workload, simplifying our migration algorithm design.

\stitle{Problem Formulation.}
Let $\mathcal{I}$ denote the set of decode instances, and for each instance $i \in \mathcal{I}$, let $B_i$ denote the set of requests assigned to instance $i$. For a request $r \in B_i$, $N(r)$ denotes the current number of tokens in request $r$, and $\hat{N}(r)$ denotes the predicted remaining generation length for request $r$. The current token load of instance $i$ is $N_i(B_i) = \sum_{r \in B_i} N(r)$.

We notice that for scheduling purposes of workload balancing, the absolute execution time is less important than the relative differences between instances. Therefore, our migration objective is to minimize the variance of token loads across all instances, which directly correlates with balancing execution time and memory usage. Variance is practical here because it supports efficient online updates and places a stronger penalty on cross-instance imbalance.

The current variance of token loads across all instances is:

\begin{equation}
    \sigma^2_0 = \mathrm{Var}(\{ N_i(B_i) \})
\end{equation}
To capture the dynamic nature of workloads where requests continuously generate new tokens or complete, we formulate our objective as minimizing the expected variance over a prediction horizon. Let $\hat{N}_i(B_{i,t})$ denote the predicted token load of instance $i$ at future time step $t$, computed using our generation length predictor. The expected variance combines both current and predicted states:

\begin{equation}
    \hat{\sigma}^2 = \sigma^2_0 + \sum_{t=1}^{\infty} \beta_t \cdot \mathrm{Var}(\{ \hat{N}_i(B_{i,t}) \}),
\end{equation}
where $\beta_t$ is a time-dependent weighting factor that balances the importance of current state versus predicted future states at time step $t$. This formulation ensures that migration decisions consider both immediate and long-term load balancing effects.

\stitle{Algorithm Design.}
The algorithm consists of two main components: \emph{scheduler-side} and \emph{worker-side} functions. The scheduler periodically collects pre-simulated state reports from all workers, classifies instances, enumerates candidates, and completes the remaining simulation to select the best feasible migration. Workers continuously monitor their local state, retrieve prediction results, perform local future state simulation, and proactively report this pre-computed information to the scheduler. STAR is a periodic online heuristic that balances execution imbalance, memory safety, and migration overhead.

Figure~\ref{fig:workflow} illustrates scheduler side workflow for our decode rescheduling architecture, which operates through three distinct phases implemented in Algorithm~\ref{alg:migration_algorithm}:

\begin{figure}[t]
    \vspace{-0.12in}
    \centering
    \includegraphics[width=1.0\columnwidth]{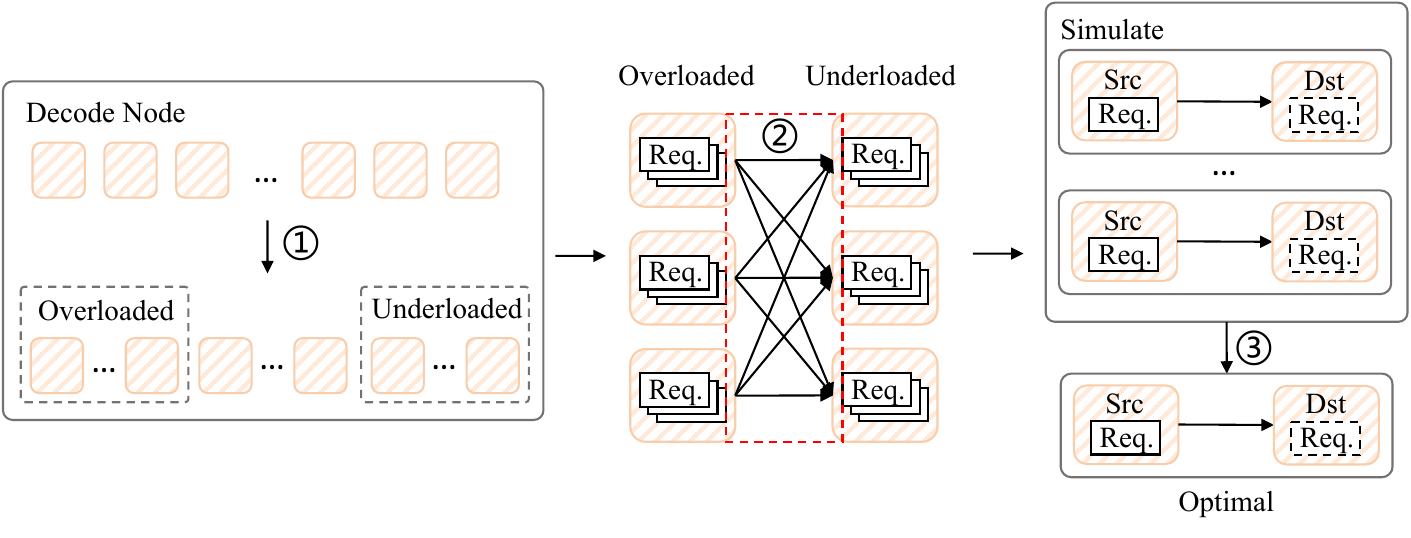}
    \caption{Workflow of the scheduler.}
    \label{fig:workflow}
\end{figure}
\begin{itemize}
    \item \textbf{Phase 1: Instance Classification (Lines 11-16).} The scheduler identifies overloaded and underloaded instances by computing the weighted workload $w_i$ for each instance and comparing it against the average workload $\bar{w}$. This classification, implemented in Lines 14-16, focuses the algorithm on a small subset of instances that require attention.

    \item \textbf{Phase 2: Candidate Enumeration (Lines 17-23).} For each source-target instance pair, the algorithm enumerates migration candidates by filtering requests that satisfy two constraints: (1) sufficient remaining tokens exceeding migration overhead (Line 20), and (2) memory safety ensuring no OOM on target instance in the near future (Line 21).

    \item \textbf{Phase 3: Best-Feasible Selection (Lines 24-34).} The algorithm evaluates each candidate by aggregating pre-computed worker simulations (Line 29) and computing the incremental variance reduction (Line 30). When prediction is enabled, the algorithm simulates future variance based on predicted token loads; otherwise, it evaluates candidates based on the current variance. The best feasible migration is selected as the one that maximizes variance reduction, leveraging the fact that workers have already performed local simulations to minimize scheduler computation overhead. Choose the migration that yields the greatest reduction in time-weighted token load variance and then execute migration.

\end{itemize}

\begin{algorithm}[htbp]
    \caption{Decode Rescheduling}
    \label{alg:migration_algorithm}
    \begin{algorithmic}[1]
        \STATE \textbf{Scheduler loop}
        \WHILE{serving requests}
            \STATE Wait for the next scheduling interval
            \STATE Collect worker states and update the average workload $\bar{w}$
            \STATE $(\mathcal{O}, \mathcal{U}) \gets \textsc{InstanceClassification}(\bar{w})$
            \IF{$\mathcal{O} \neq \emptyset$}
                \STATE $\mathcal{C} \gets \textsc{CandidateEnumeration}(\mathcal{O}, \mathcal{U}, \mathcal{S})$
                \STATE $m^* \gets \textsc{BestFeasibleSelection}(\mathcal{C}, \sigma^2_0)$
                \IF{$m^* \neq \emptyset$}
                    \STATE $\textsc{ExecuteMigration}(m^*)$
                \ENDIF
            \ENDIF
        \ENDWHILE

        \STATE \textbf{Function} \textsc{InstanceClassification}($\bar{w}$)
        \FOR{each instance $i \in \mathcal{I}$}
            \STATE $w_i = \sum_{t=1}^{H} \beta_t \cdot \hat{N}_i(B_{i,t})$
        \ENDFOR
        \STATE $\mathcal{O} \gets \{i \in \mathcal{I} \mid w_i > (1+\theta) \cdot \bar{w}\}$
        \STATE $\mathcal{U} \gets \{i \in \mathcal{I} \mid N_i(B_{i,0}) < (1+\theta) \cdot \bar{w}\}$
        \STATE \textbf{return} $(\mathcal{O}, \mathcal{U})$

        \STATE \textbf{Function} \textsc{CandidateEnumeration}($\mathcal{O}, \mathcal{U}, \mathcal{S}$)
        \STATE $\mathcal{C} \gets \emptyset$
        \FOR{each $(s, t) \in \mathcal{O} \times \mathcal{U}$}
            \STATE $\mathcal{R}_s \gets \{r \in B_s \mid \hat{N}(r) > \frac{C_{\text{mig}}}{\bar{T}_{\text{exec}}}\}$
            \STATE $\mathcal{R}_s \gets \mathcal{R}_s \cap \{r \mid N_t(B_{t,0}) + \hat{N}(r) \leq C_{\text{mem}}\}$
            \STATE $\mathcal{C} \gets \mathcal{C} \cup \{(r, s, t) \mid r \in \mathcal{R}_s\}$
        \ENDFOR
        \STATE \textbf{return} $\mathcal{C}$

        \STATE \textbf{Function} \textsc{BestFeasibleSelection}($\mathcal{C}, \sigma^2_0$)
        \STATE $m^* \gets \text{null}$
        \STATE $\sigma^2_{\max} \gets 0$
        \FOR{each candidate $(r, s, t) \in \mathcal{C}$}
            \IF{$\text{usePrediction}$}
                \STATE $\hat{\mathcal{S}} \gets \text{AggregateSimulations}(\mathcal{S}, r, s, t)$
                \STATE $\sigma^2 \gets \text{SimulateFutureVariance}(\hat{\mathcal{S}}, H)$
            \ELSE
                \STATE $\sigma^2 \gets \text{CurrentVariance}(\mathcal{S}, r, s, t)$
            \ENDIF
            \STATE $m^* \gets \text{UpdateBestCandidate}(m^*, \sigma^2, \sigma^2_{\max})$
        \ENDFOR
        \STATE \textbf{return} $m^*$
    \end{algorithmic}
\end{algorithm}

\stitle{Complexity Analysis.}
Let $n=|\mathcal{I}|$ be the number of decode instances, let $\mathcal{O}$ and $\mathcal{U}$ be the overloaded and underloaded instance sets, let $R_{\max}$ be the maximum number of active requests on any instance, and let $H$ be the prediction horizon. In the native implementation of Algorithm~\ref{alg:migration_algorithm}, instance classification costs $O(n)$, and candidate enumeration costs $O(|\mathcal{O}| \cdot |\mathcal{U}| \cdot R_{\max})$. For each source-target pair, there are at most $R_{\max}$ candidate requests, and evaluating one candidate requires recomputing an $H$-step token-load trace over up to $R_{\max}$ requests, which costs $O(R_{\max}H)$. Therefore, the native scheduler complexity per interval is $O(n + |\mathcal{O}| \cdot |\mathcal{U}| \cdot R_{\max}^2 \cdot H)$. This cost can be reduced with worker-side pre-aggregation and scheduler-side incremental updates. Each worker first computes its $H$-step future-load summary once in $O(R_{\max}H)$ time. Then each candidate evaluation only updates the source and target summaries, reducing the cost to $O(H)$. The optimized scheduler complexity becomes $O(n + |\mathcal{O}| \cdot |\mathcal{U}| \cdot R_{\max} \cdot H)$. In Section~\ref{subsec:large_scale_simulation}, these computations remain below 300\,ms even for 256 instances and can overlap with decode computation.

\subsection{Trade-off between Frequency and Overhead}
We notice that frequent reprediction and rescheduling, e.g., at every decode iteration, can improve the precision of the predicted remaining generation length as well as the workload balancing effectiveness. However, this comes at the cost of increased computational overhead incurred by the prediction model.

Specifically, consider the model DeepSeek-R1-Distill-Qwen-7B tested on RTX 4090D, our prediction model takes 1.40 ms to infer a batch of 10 requests, while each decode iteration takes about 18.23\,ms under 50\% KV cache memory occupancy. Therefore, injecting prediction at every decode iteration would incur a prohibitive overhead of 7.68\%, which is unacceptable in practice.
Formally, setting prediction interval to every $k$ decode iterations results in a prediction overhead of $\frac{1.40}{18.23 \times k}$.
Regarding workload characteristics, $k$ iterations at most increases the workload by $\frac{k}{l}$, where $l$ represents the average length of requests in decode instances (e.g., 2000 tokens in our settings).
Considering both prediction accuracy and computational overhead, we can set $k$ to 20 as the prediction interval, which only incurs a low overhead of 0.38\% while maintaining high prediction accuracy with less than 1\% error.

\subsection{Overlap Migration with Decode Computation}
We notice that the migration of decode requests inevitably incurs overhead due to KV cache transfer, which can overlap with decode computation to minimize its impact on overall system performance.
Specifically, migration process leverages NVIDIA Inference Transfer Library for asynchronous KV cache transfer. The paused request's KV cache is transferred to the target instance without blocking the execution of other requests in the same batch. This asynchronous design ensures that migration overhead does not impact the performance of non-migrating requests.
To ensure seamless user experience, we implement a proxy-based architecture where users establish connections with proxy that is decoupled from the processing instances. The proxy maintains a persistent stream connection with the client, enabling continuous response delivery during request migration, ensuring that users remain unaware of migration events.

\section{Evaluation}\label{sec:evaluation}
\subsection{Experimental Setup}
\label{subsec:setup}
\stitle{Implementation.}
We implement \ours in approximately 6,000 lines of code (LoC) on top of vLLM 0.9.2, which supports PD disaggregation with the official NIXL (NVIDIA Inference Transfer Library) component for KV cache transfers.
\ours includes three main components: 1) the generation length predictor (Section~\ref{sec:prediction}), 2) the rescheduling algorithm (Section~\ref{sec:scheduler}), and 3) the migration proxy that orchestrates migrations between decode instances.
For the predictor, we implement a custom MLP architecture and develop training and evaluation scripts that integrate seamlessly into the vLLM engine. The rescheduling algorithm is implemented as a standalone module that monitors decode instances and executes migration decision process. The migration proxy extends the vLLM engine with decode-to-decode KV cache transfer capabilities and runs rescheduling algorithm.
Additionally, we develop a dedicated simulator to validate the theoretical effectiveness of our approach in large-scale systems, which will be detailed in Section~\ref{subsec:large_scale_simulation}.

\begin{figure*}[t]
    \centering
    \subfigure[Goodput on ShareGPT (small cluster).]{
        \includegraphics[width=0.30\textwidth]{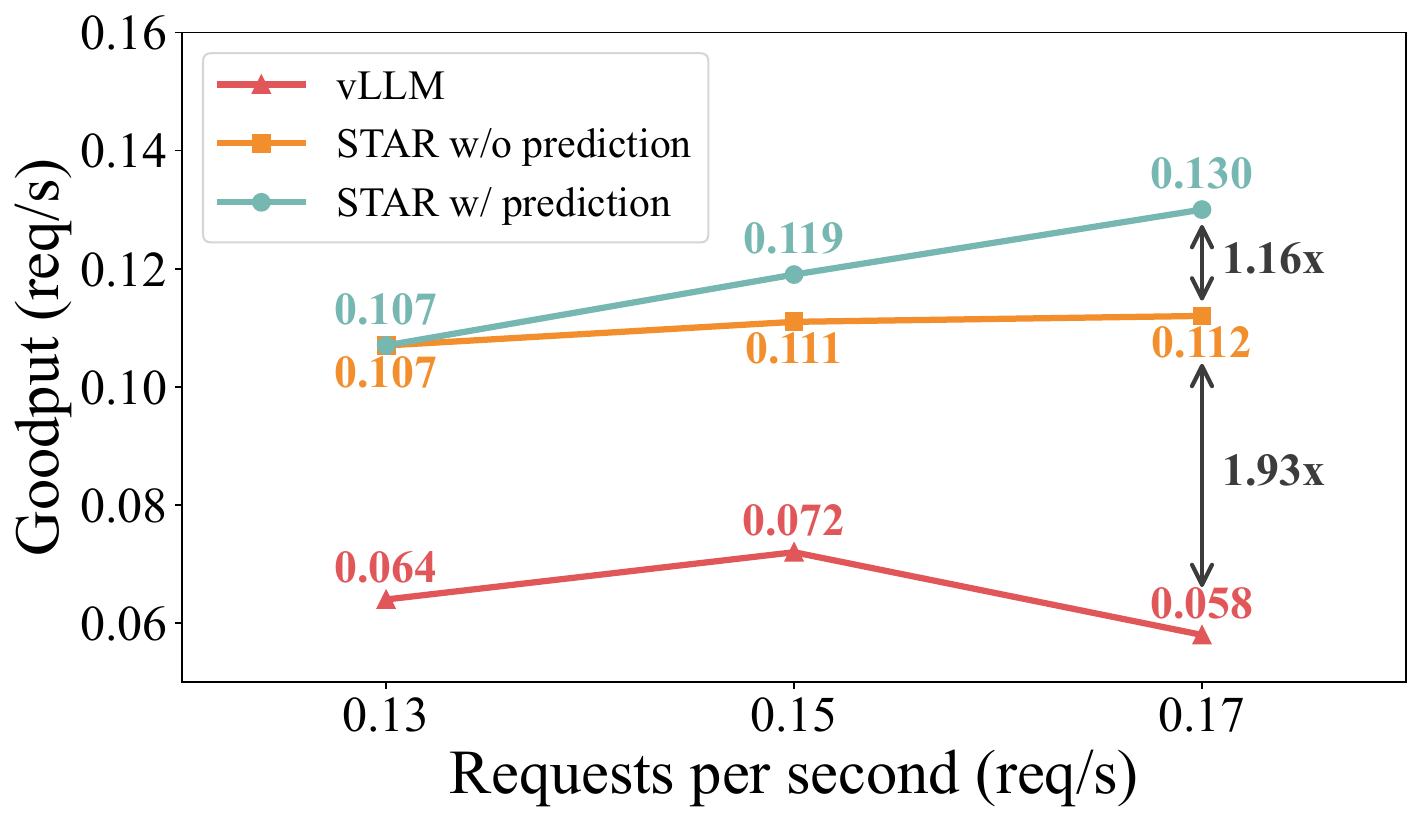}
        \label{fig:qps_goodput}
    }
    \hfill
    \subfigure[Throughput on ShareGPT (small cluster).]{
        \includegraphics[width=0.30\textwidth]{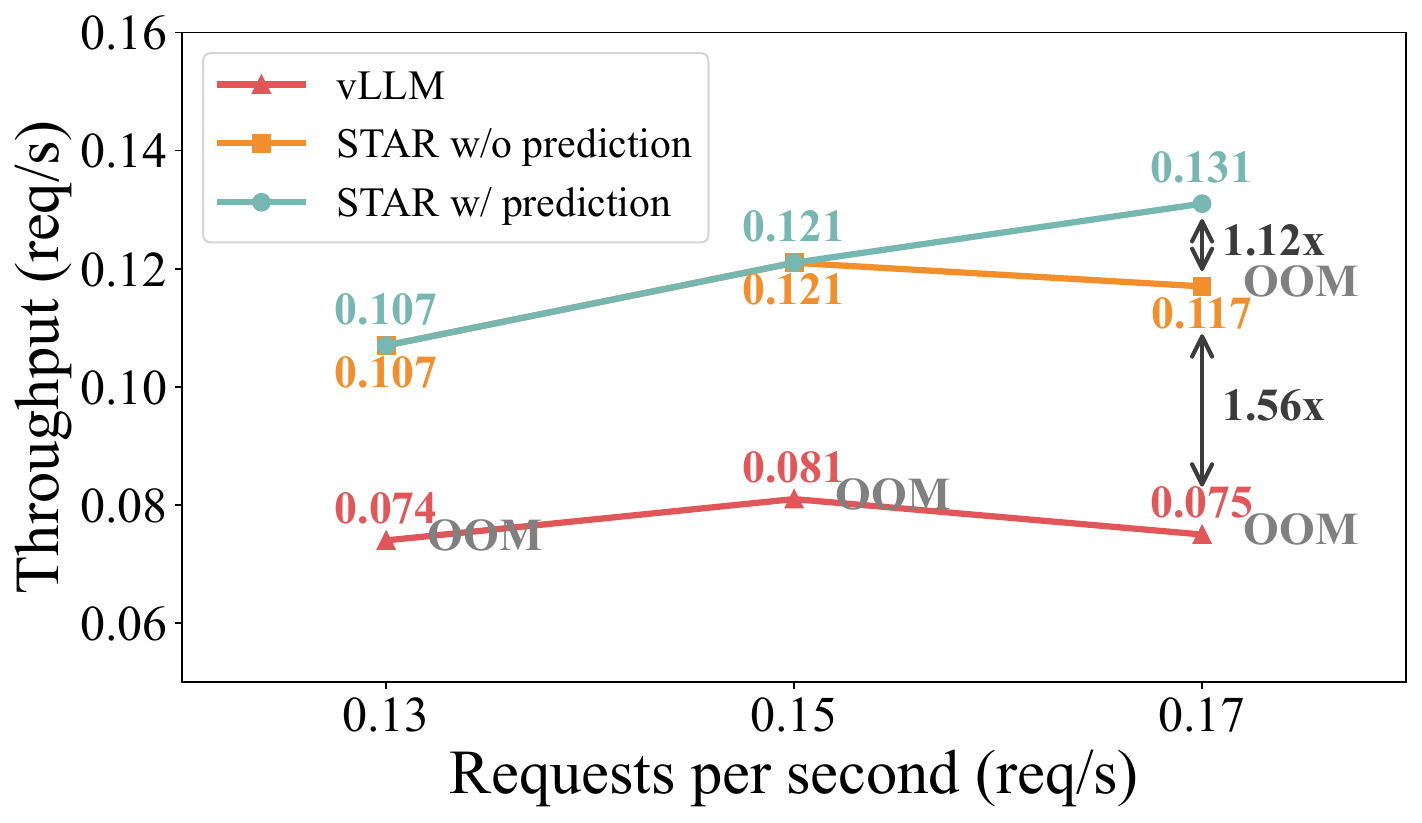}
        \label{fig:throughput_req_qps}
    }
    \hfill
    \subfigure[P99 latency on ShareGPT (small cluster).]{
        \includegraphics[width=0.29\textwidth]{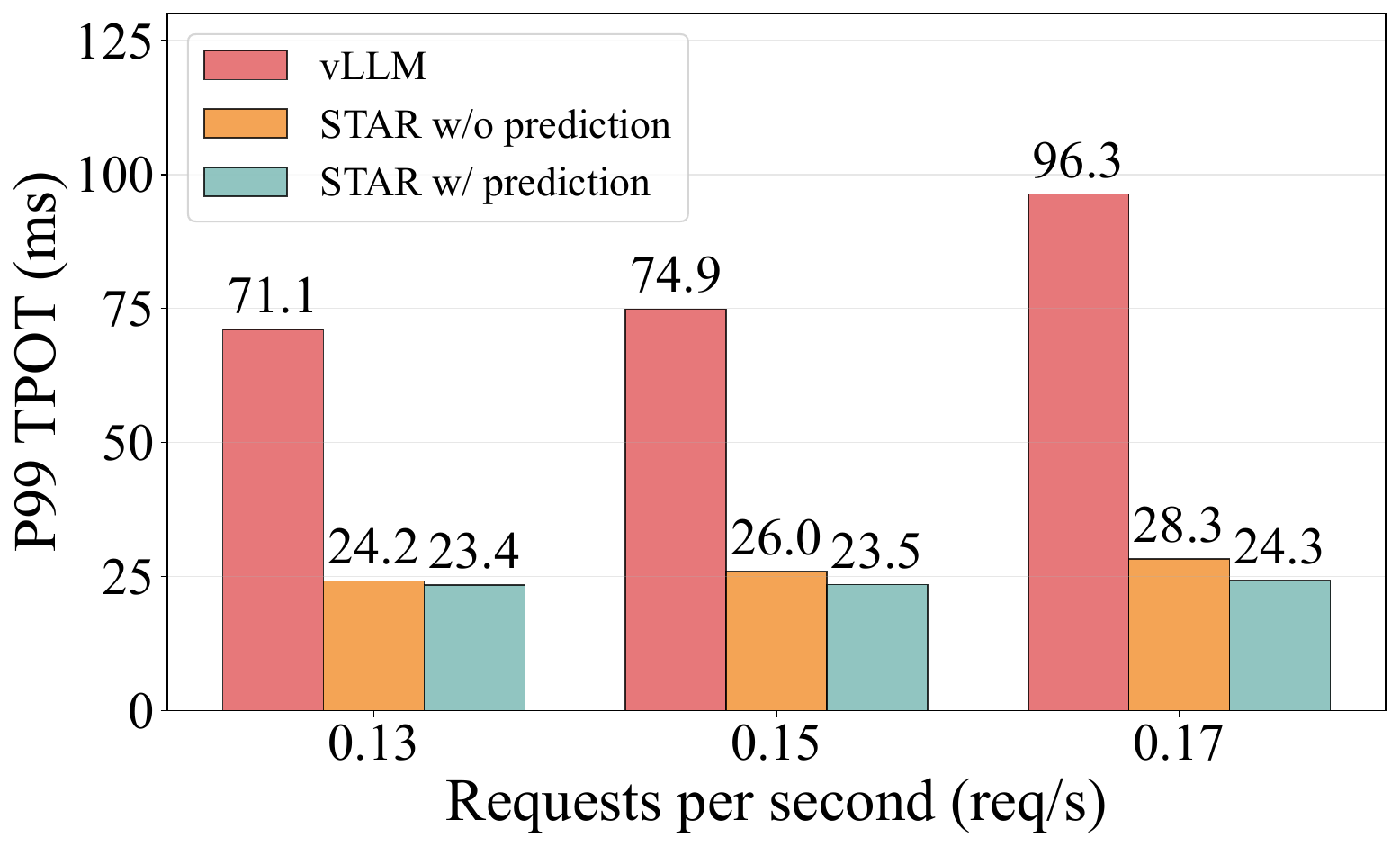}
        \label{fig:sharegpt_p99_qps_comparison}
    }
    \\[-5pt]
    \subfigure[Goodput on Alpaca (small cluster).]{
        \includegraphics[width=0.30\textwidth]{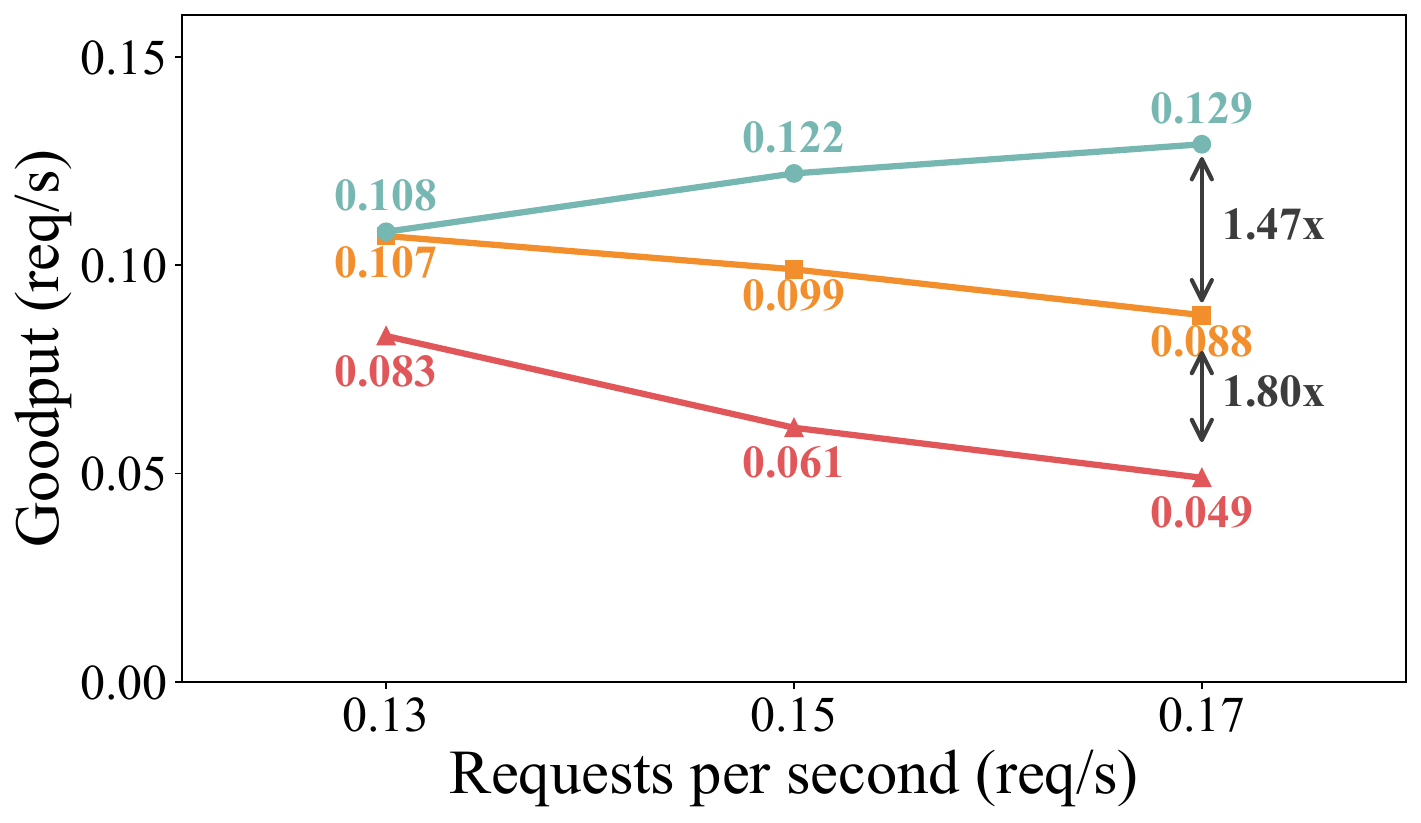}
        \label{fig:qps_goodput_alpaca}
    }
    \hfill
    \subfigure[Throughput on Alpaca (small cluster).]{
        \includegraphics[width=0.30\textwidth]{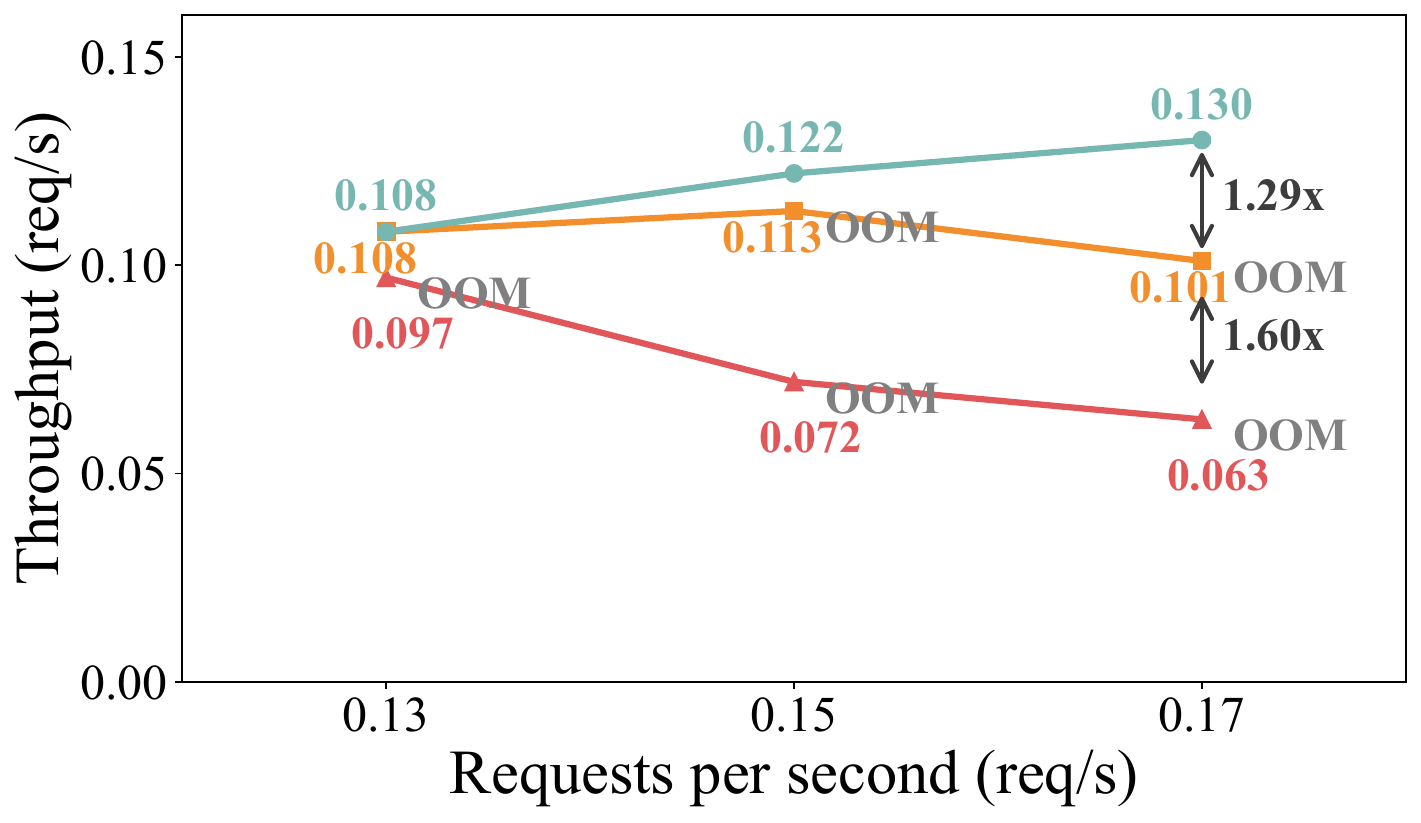}
        \label{fig:throughput_req_qps_alpaca}
    }
    \hfill
    \subfigure[P99 latency on Alpaca (small cluster).]{
        \includegraphics[width=0.29\textwidth]{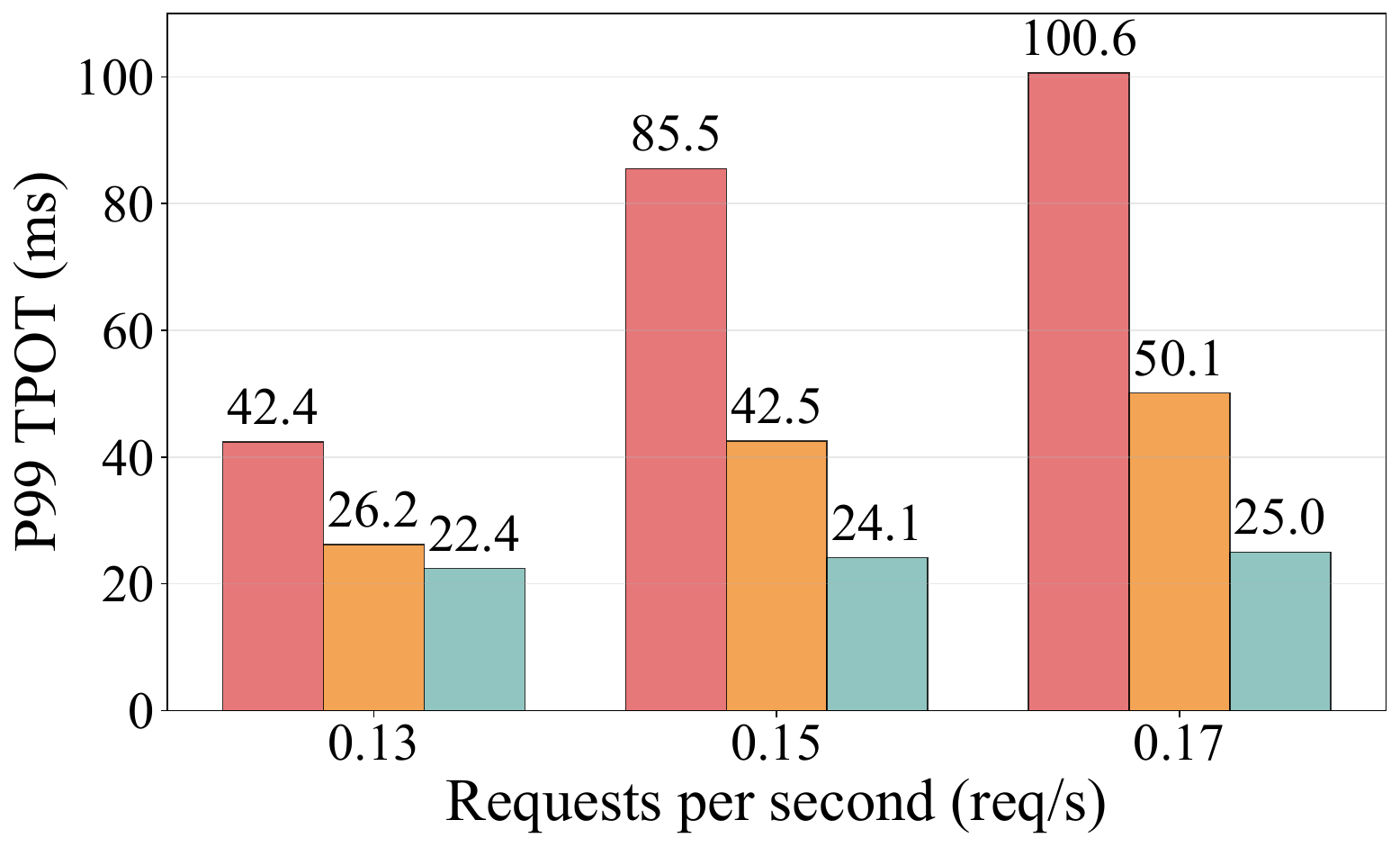}
        \label{fig:alpaca_p99_qps_comparison}
    }
    \\[-2pt]
    \subfigure[Goodput on ShareGPT (large cluster).]{
        \includegraphics[width=0.30\textwidth]{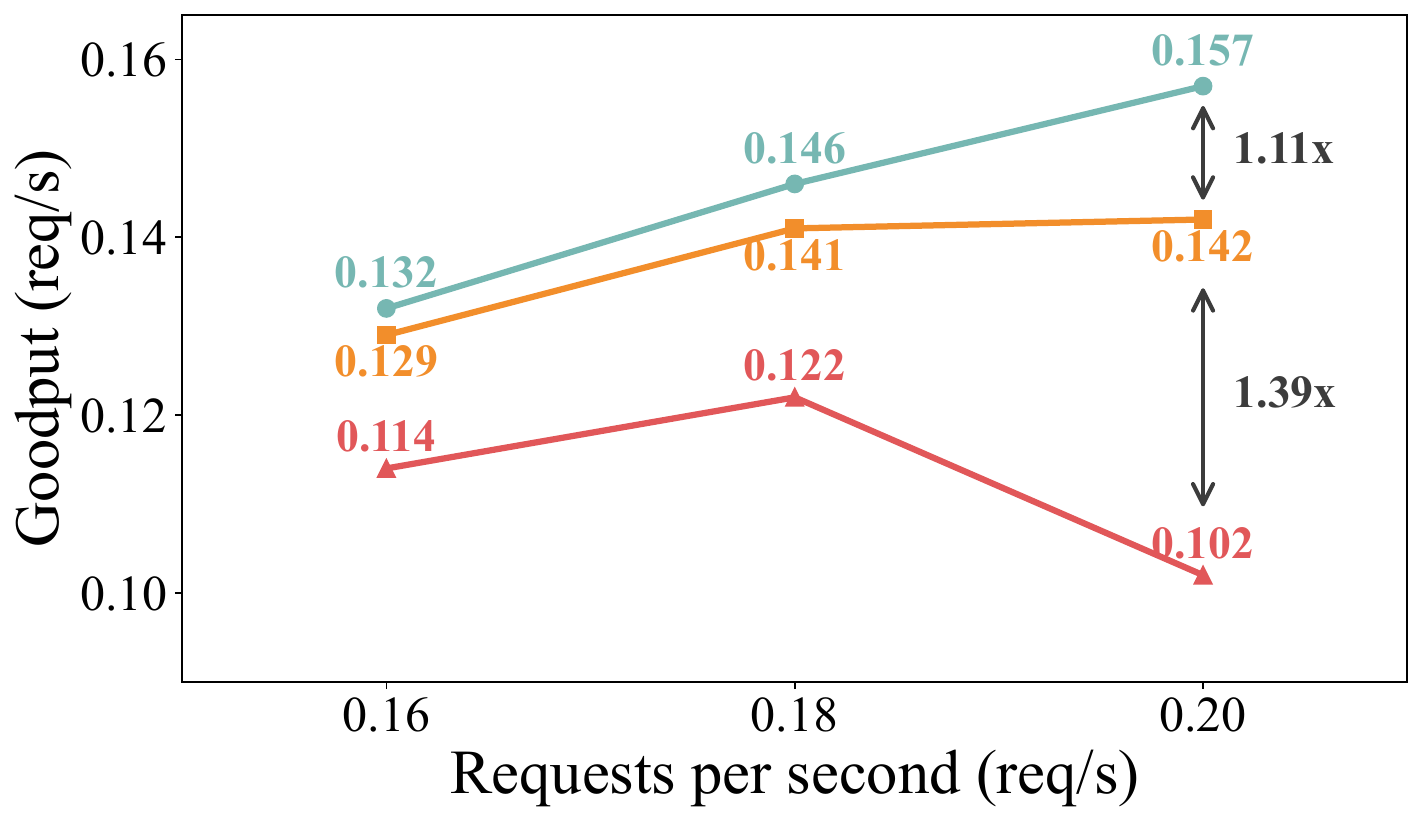}
        \label{fig:qps_goodput_sharegpt_h800}
    }
    \hfill
    \subfigure[Throughput on ShareGPT (large cluster).]{
        \includegraphics[width=0.30\textwidth]{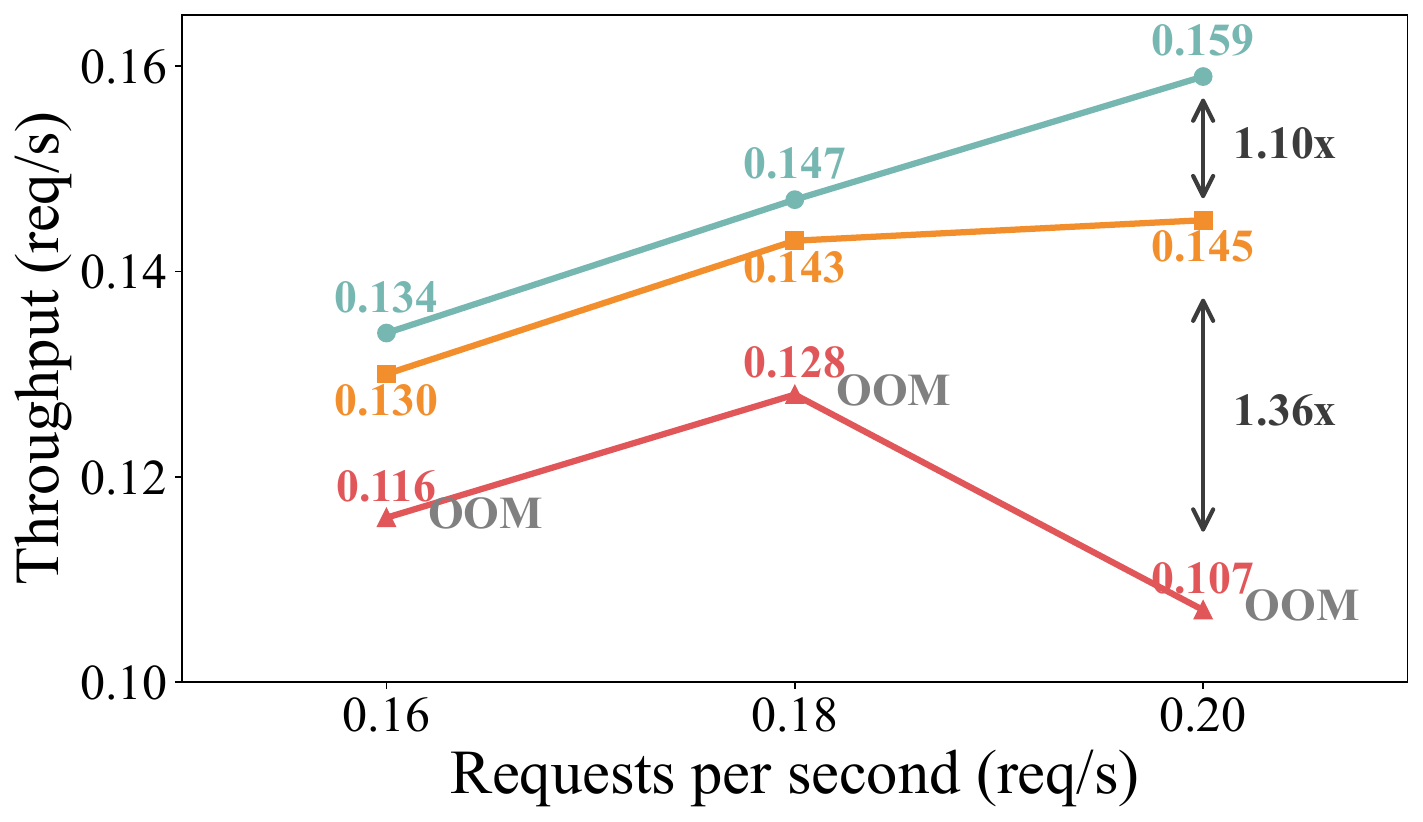}
        \label{fig:throughput_req_qps_sharegpt_h800}
    }
    \hfill
    \subfigure[P99 latency on ShareGPT (large cluster).]{
        \includegraphics[width=0.29\textwidth]{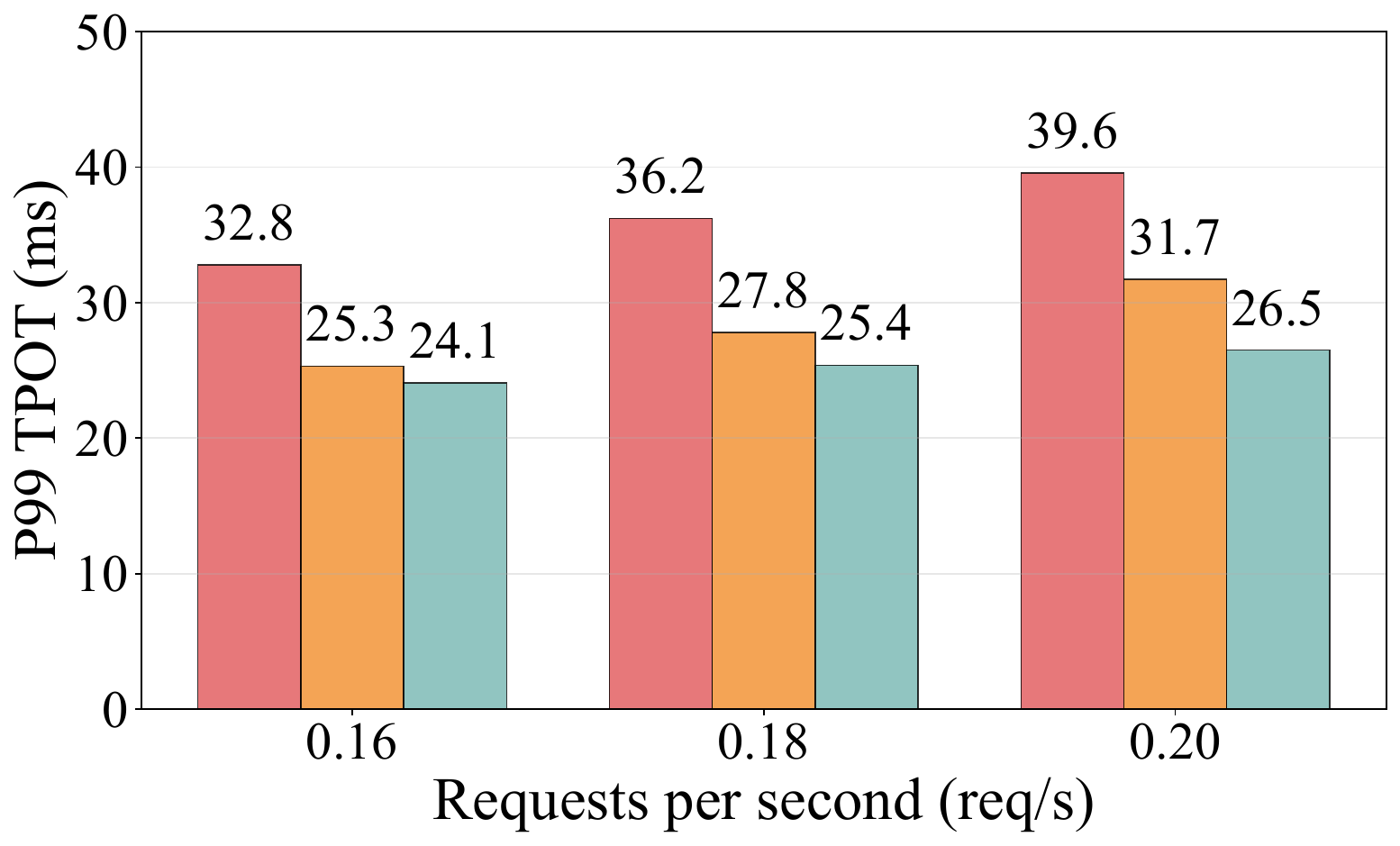}
        \label{fig:sharegpt_p99_qps_comparison_h800}
    }
    \vspace{-2pt}
    \caption{Overall performance on ShareGPT and Alpaca datasets.}
    \label{fig:overall_performance}
\end{figure*}

\stitle{Environment and models.}
We validate \ours across multiple hardware configurations as below:
\begin{itemize}
    \item small cluster: A server equipped with 4 NVIDIA RTX 4090D GPUs (1 for prefill and 3 for decode), a 40-core Intel Xeon Gold 5418Y CPU and 120GB memory, running Ubuntu 22.04.4 LTS and CUDA 12.2.
    \item large cluster: A server equipped with 8 NVIDIA H800 GPUs (2 for prefill and 6 for decode), a 40-core Intel Xeon Platinum 8458P CPU and 400GB memory, running Ubuntu 22.04.4 LTS and CUDA 12.8.
\end{itemize}
In addition, we also develop a simulation framework to evaluate larger scale clusters as detailed in Section~\ref{subsec:large_scale_simulation}.
In order to verify the effect under long output requests, we use the reasoning model DeepSeek-R1-Distill-Qwen-7B~\cite{deepseekai2025deepseekr1incentivizingreasoningcapability} and Qwen3-14B~\cite{yang2025qwen3technicalreport} as base model and adopt W8A8 quantization (only in small cluster) to support longer output tokens~\cite{frantar2023gptqaccurateposttrainingquantization}, which supports up to 32K tokens for both prompt and output.
\begin{table}[tbp]
    \centering
    \caption{Workload Statistics}
    \label{tab:workload_stats}
    \begin{tabular}{@{}l|cccccc@{}}
        \toprule
        Workload                  & Metric & Mean & Std   & P50  & P90   & P95   \\
        \midrule
        \multirow{2}{*}{ShareGPT} & Input  & 305  & 1053  & 36   & 920   & 1609  \\
                                  & Output & 7542 & 12008 & 1536 & 32670 & 32679 \\
        \midrule
        \multirow{2}{*}{Alpaca}   & Input  & 11   & 4     & 10   & 15    & 18    \\
                                  & Output & 8596 & 13354 & 987  & 32690 & 32691 \\
        \bottomrule
    \end{tabular}
\end{table}

\stitle{Workload.}
We utilize two widely adopted real-world datasets, ShareGPT~\cite{sharegpt} and Alpaca~\cite{alpaca}, to evaluate the performance of \ours.
By default, we present results using the ShareGPT dataset.
The prompt and generation length distribution of the dataset using DeepSeek-R1-Distill-Qwen-7B are shown in Table~\ref{tab:workload_stats}, where P50, P90, and P95 represent the 50th, 90th, and 95th percentiles of the length distribution starting from the shortest request. Notably, over 15\% of requests generate over 25K tokens, highlighting the challenge of load balancing for decode instances.

\stitle{Baselines.}
We evaluate the following migration strategies with \textit{real-time KV cache load balancing} for the prefill-to-decode transfer (serving as the common foundation):
\begin{itemize}
    \item \textit{vLLM~\cite{kwon2023efficient}}: We employ vLLM's built-in prefill-decode disaggregation architecture as the baseline, which adopts the idea from DistServe~\cite{zhong2024distserve}. As the industry-standard serving framework, \textit{vLLM} already supports PD disaggregation.
    \item \textit{\ours w/o prediction}: We implement a rescheduling algorithm without prediction upon the vLLM framework, which periodically evaluates the workload of decode instances and migrates requests from overloaded to underloaded instances based on current state only.
    \item \textit{\ours w/ prediction}: We integrate the generation length predictor and the decode rescheduling algorithm.
    \item \textit{\ours Oracle}: An oracle version of \textit{{\ours w/ prediction}} that assumes exact remaining generation lengths for all active requests, providing an upper bound on prediction-aware rescheduling.
\end{itemize}

\subsection{End-to-End Performance}

Figure~\ref{fig:overall_performance} illustrates the three critical metrics: throughput, and goodput, and tail latency, across different scheduling strategies under various requests per second (RPS). Subsequently, we analyze each metric in detail.

\stitle{Throughput.}
As shown in Figure~\ref{fig:throughput_req_qps}, Figure~\ref{fig:throughput_req_qps_alpaca} and Figure~\ref{fig:throughput_req_qps_sharegpt_h800}, both rescheduling and prediction improve throughput, with the largest gain under high load. On large cluster at 0.20 req/s, rescheduling raises throughput from 0.107 to 0.145 req/s (35.5\%), and prediction further raises it to 0.159 req/s (an additional 9.7\%). This gain comes from alleviating decode-instance saturation and OOM caused by workload imbalance.

\stitle{Goodput.}
We use goodput to measure the requests that satisfy the SLO. The SLO is 1s TTFT, with TPOT targets of 25\,ms for DeepSeek-R1-Distill-Qwen-7B and 50\,ms for Qwen3-14B. Because workload imbalance directly increases tail latency and SLO violations, goodput shows the benefit of rescheduling and prediction more clearly than throughput.
Figure~\ref{fig:qps_goodput_sharegpt_h800} shows an even larger gain in goodput. On large cluster at 0.20 req/s, rescheduling improves goodput from 0.102 to 0.142 req/s (39.2\%), and prediction further improves it to 0.157 req/s (an additional 10.6\%). This indicates fewer SLO violations under better-balanced decode load.

\stitle{Latency.}
We report P99 TPOT to capture tail latency. Figure~\ref{fig:sharegpt_p99_qps_comparison} and Figure~\ref{fig:alpaca_p99_qps_comparison} show large reductions across both datasets. On ShareGPT at RPS=0.17, rescheduling reduces P99 TPOT from 96.3\,ms to 28.3\,ms, and prediction further reduces it to 24.3\,ms.
Figure~\ref{fig:sharegpt_p99_qps_comparison_h800} shows a similar trend. At 0.20 req/s, rescheduling reduces P99 TPOT from 39.57\,ms to 31.72\,ms, and prediction further reduces it to 26.49\,ms. This reduction comes from preventing heavily overloaded decode instances and the resulting tail-latency inflation.

\subsection{Scalability and Load-Balance Analysis}
\label{subsec:large_scale_simulation}

To demonstrate the performance upper bound of our prediction-based rescheduling approach, we introduce an oracle version \textit{\ours Oracle}, which assumes perfect knowledge of the generation lengths for all requests. Subsequently, we evaluate the effectiveness of our approach in both small-scale real systems and large-scale simulated clusters by demonstrating the variation in execution time across decode instances for 2,000 seconds on shareGPT dataset.

\begin{figure}[htbp]
    \centering
    \includegraphics[trim=18 12 18 18,clip,width=0.96\columnwidth]{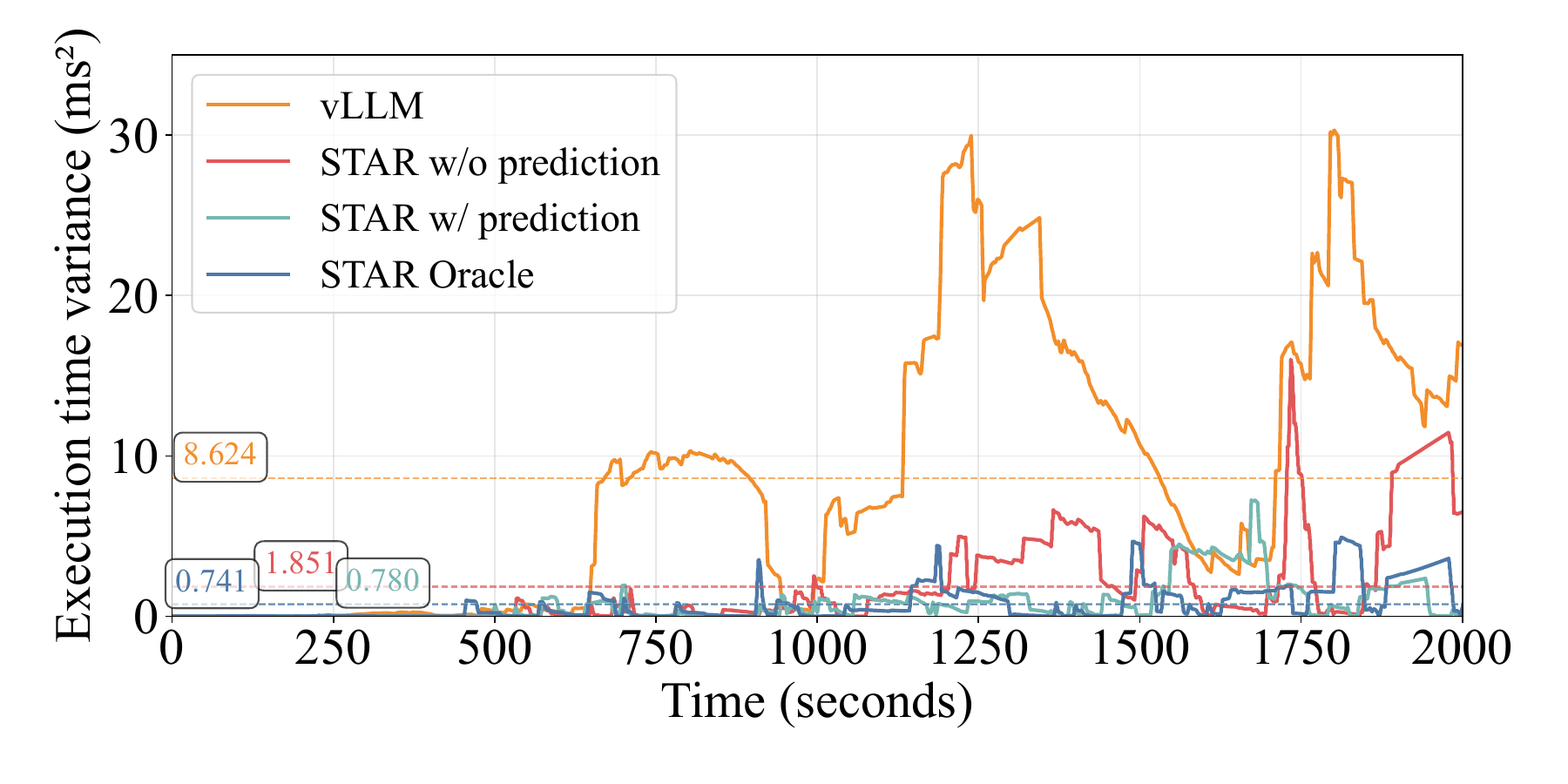}
    \caption{Execution time variance across different scheduling algorithms on high-load dataset.}
    \label{fig:tpot_variance_comparison}
    \vspace{-3pt}
\end{figure}

\stitle{Execution on small scale.}
Figure~\ref{fig:tpot_variance_comparison} illustrates the execution time variance across different scheduling strategies when running on 3 decode instances in small cluster.
Our proposed prediction solution achieved an average execution time variance of 0.78 ms², which is close to that of oracle prediction.
\textit{vLLM shows} bursty variance under workload imbalance, rescheduling mitigates it, and prediction reduces it further.

\begin{figure*}[t]
    \centering
    \subfigure[2 prefill, 6 decode.]{
        \includegraphics[width=0.32\textwidth]{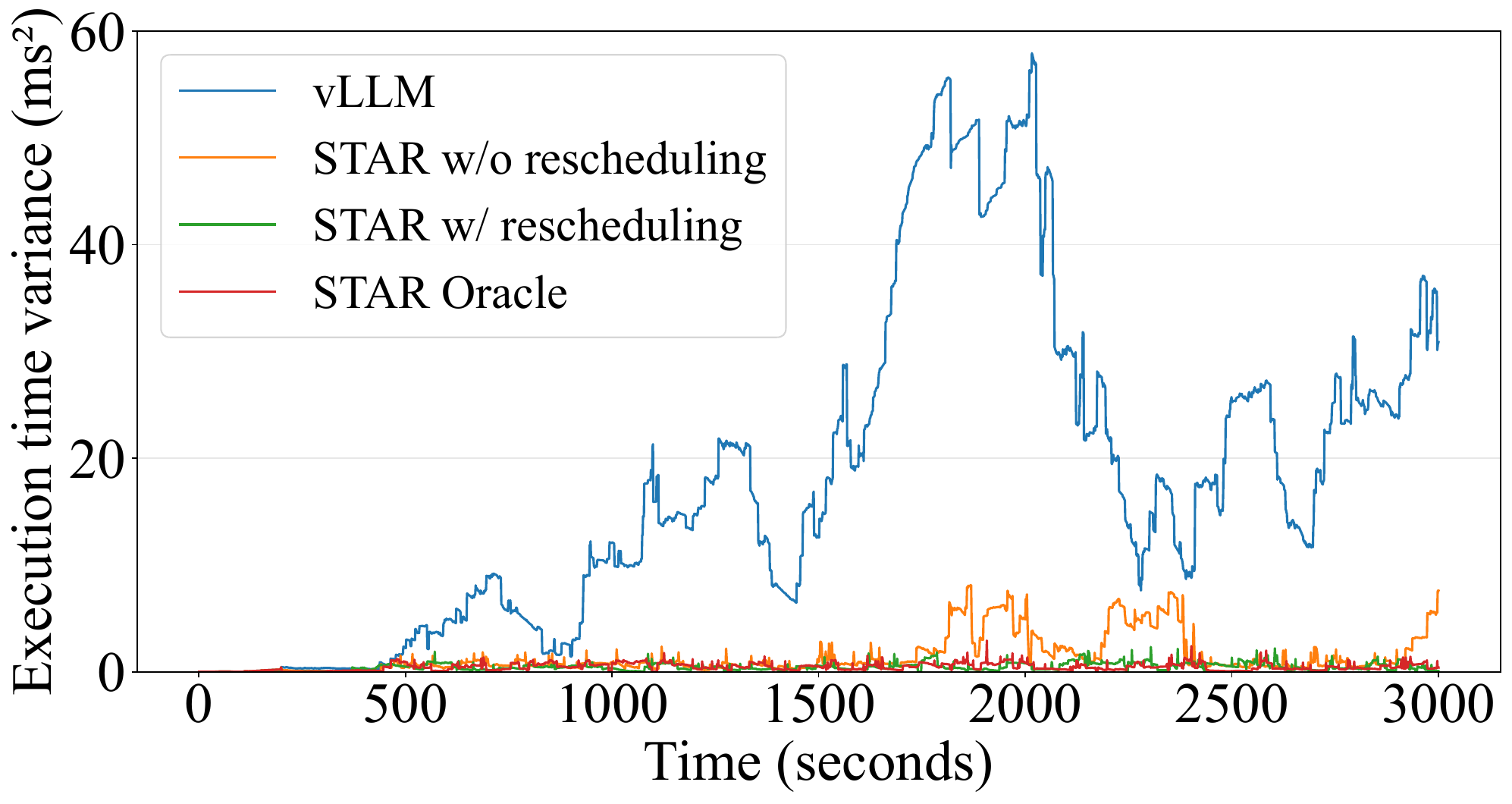}
    }
    \hfill
    \subfigure[4 prefill, 12 decode.]{
        \includegraphics[width=0.32\textwidth]{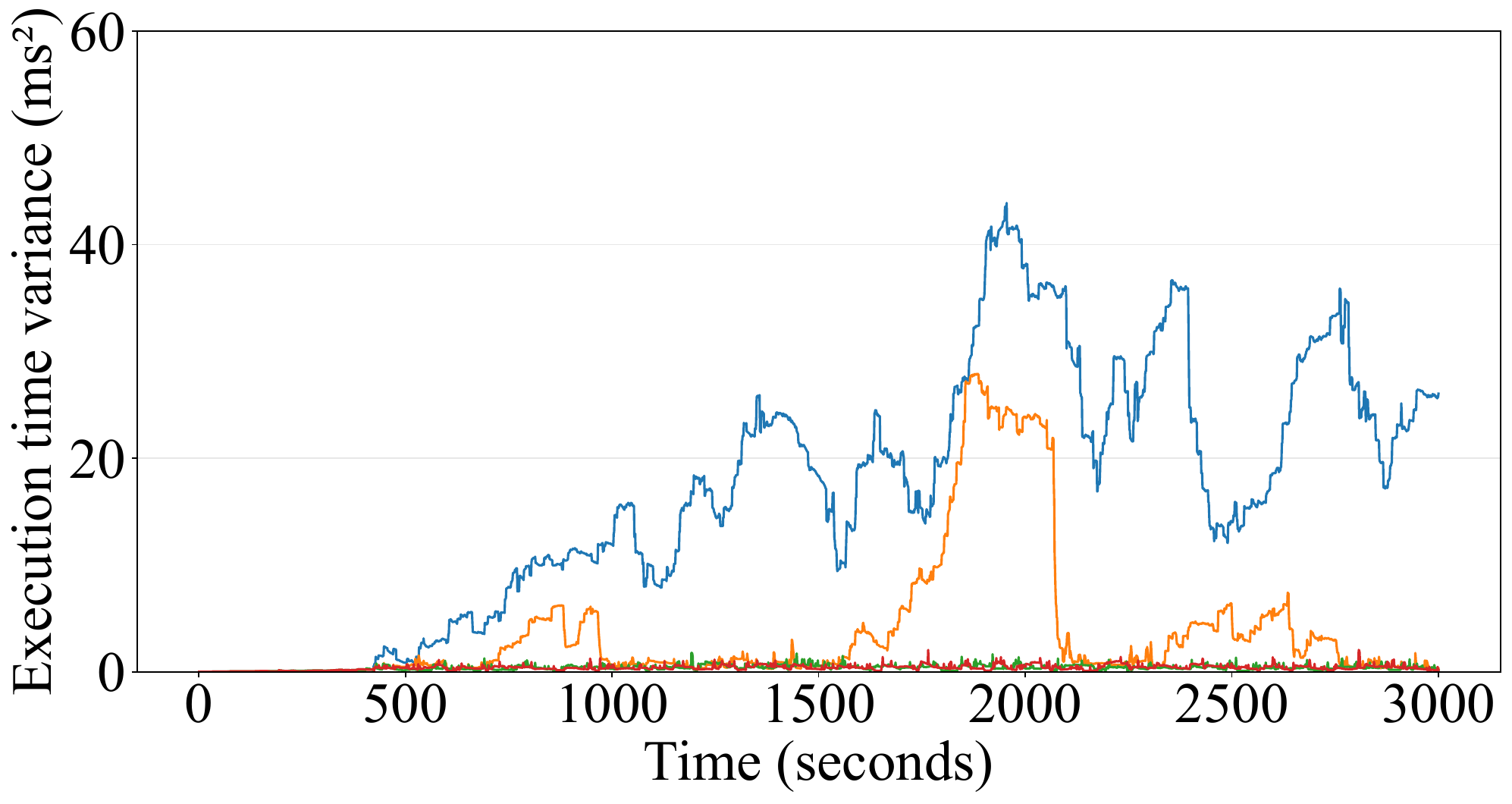}
    }
    \hfill
    \subfigure[8 prefill, 24 decode.]{
        \includegraphics[width=0.32\textwidth]{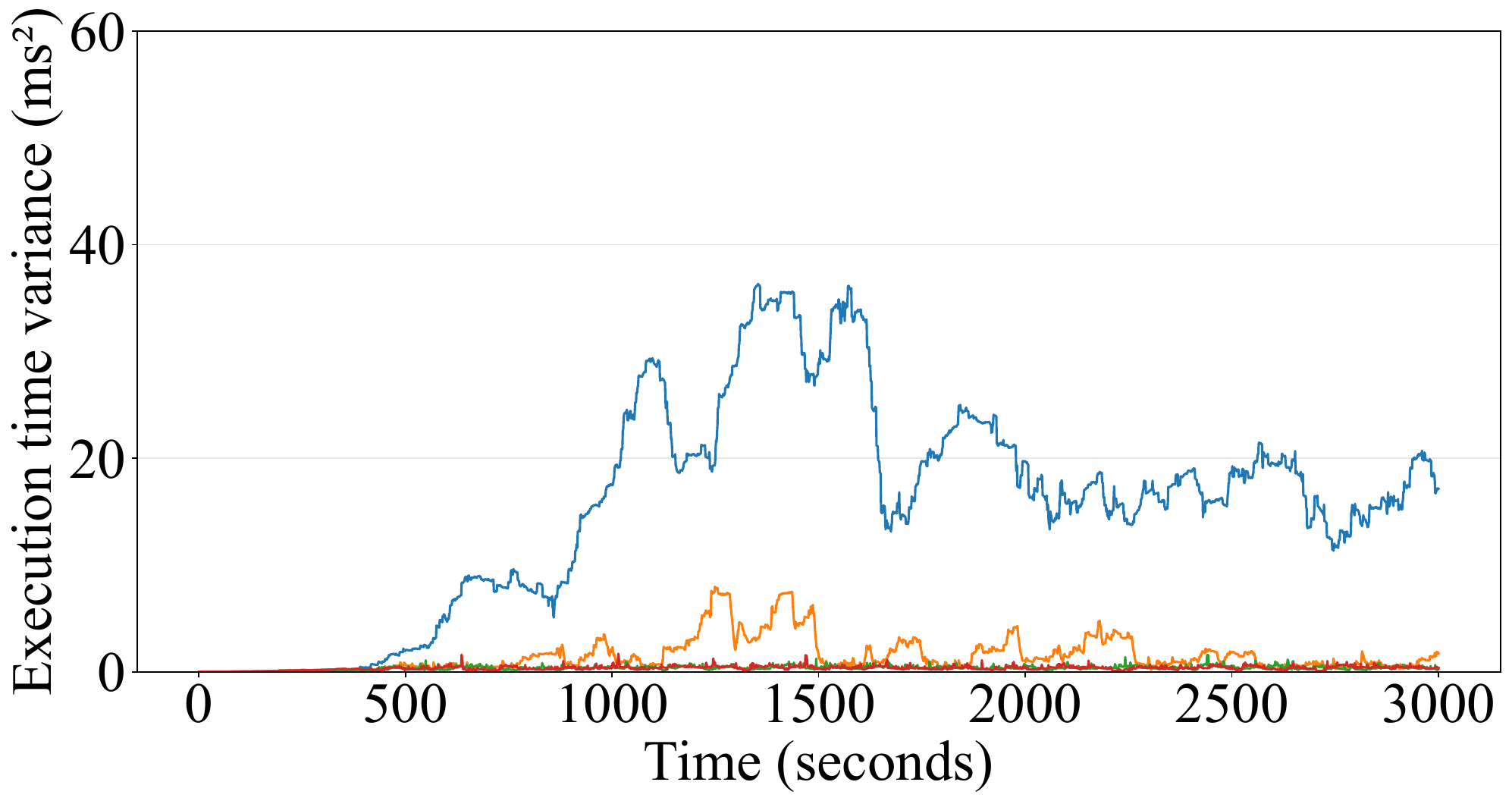}
    }
    \\[-3pt]
    \subfigure[16 prefill, 48 decode.]{
        \includegraphics[width=0.32\textwidth]{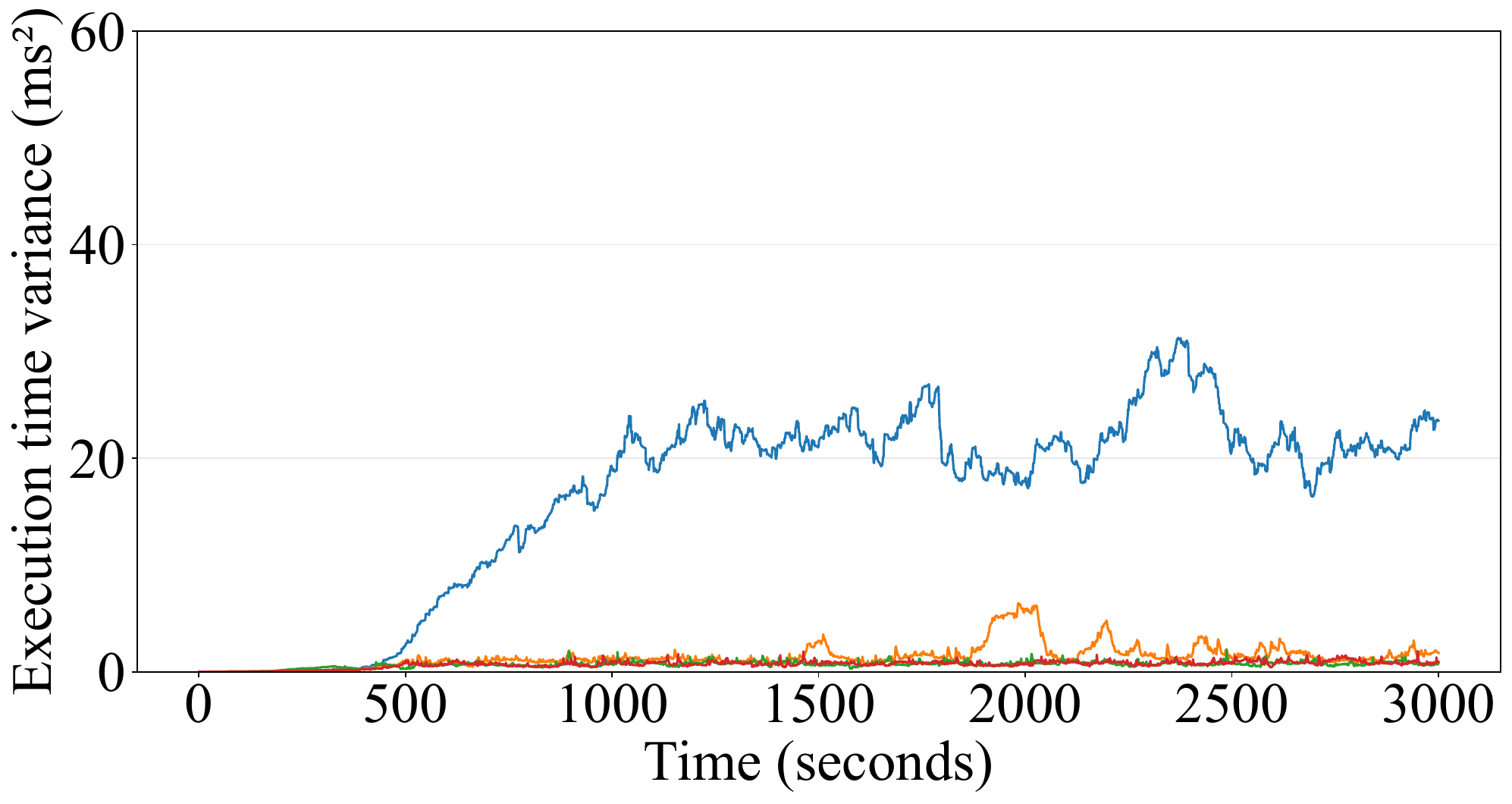}
    }
    \hfill
    \subfigure[32 prefill, 96 decode.]{
        \includegraphics[width=0.32\textwidth]{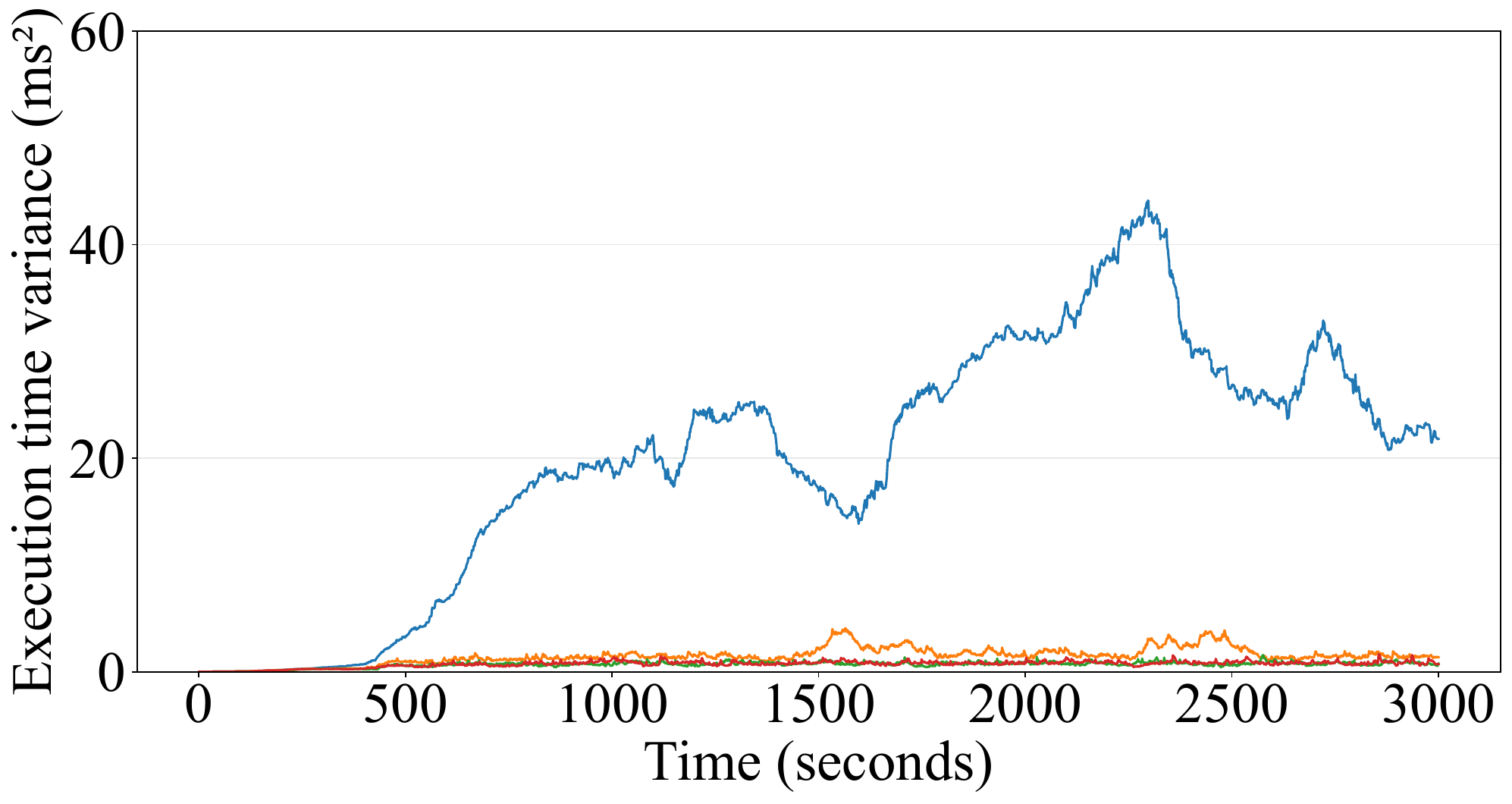}
    }
    \hfill
    \subfigure[64 prefill, 192 decode.]{
        \includegraphics[width=0.32\textwidth]{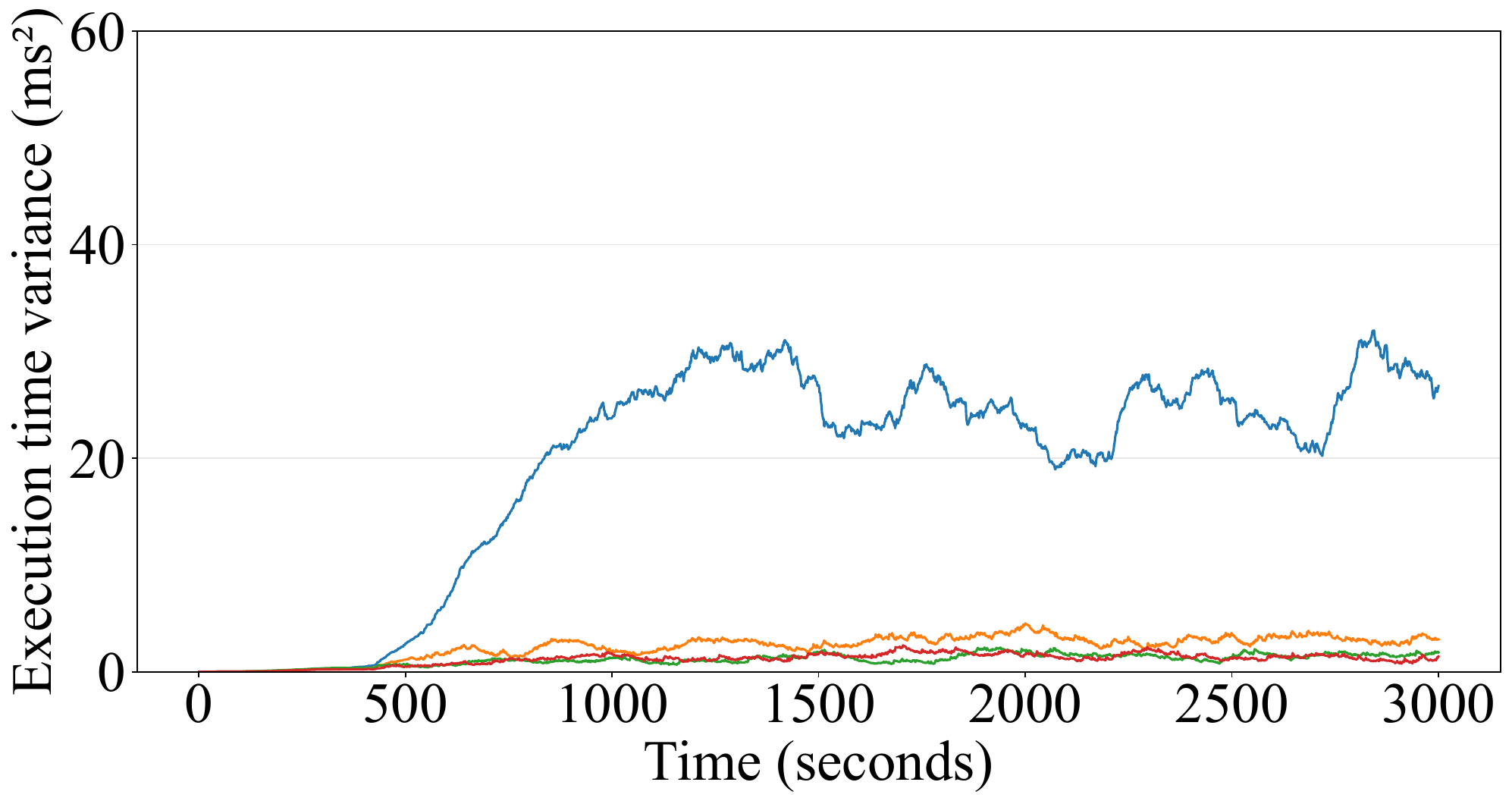}
    }
    \vspace{-2pt}
    \caption{Execution time variance comparison across different cluster sizes under 25 Gbps transfer speed.}
    \label{fig:large_scale_variance}
\end{figure*}

\stitle{Simulation on large scale.}
We conduct simulations using our dedicated simulator that models large-scale cluster behaviors across hundreds of instances in production scale deployments.
The simulator follows the same scheduling and migration logic as the real system.
Specifically, we use event-driven simulation to model request arrivals, decode execution, and migration events.
The execution time of each decode iteration is derived from real system measurements, while the migration overhead is calculated based on KV cache size and network bandwidth.
Regarding prediction, we leverage the actual remaining generation lengths to simulate an oracle predictor.
We set the request rate to 0.3 RPS for an 8-instance cluster and scale it linearly with cluster size. This configuration ensures that the system reaches a dynamic equilibrium where request arrival and completion rates are balanced, preventing unbounded queue growth while maintaining sufficient workload.
Using ShareGPT and a 25 Gbps transfer speed (following DistServe's cross-node interconnection setting~\cite{zhong2024distserve}), we evaluate cluster sizes from 8 to 256 instances.
Figure~\ref{fig:large_scale_variance} shows that rescheduling improves load balance.
\textit{\ours w/ prediction} achieves load balancing close to the oracle, and this advantage persists as the cluster scales up.

\begin{figure}[htbp]
    \centering
    \includegraphics[width=0.9\columnwidth]{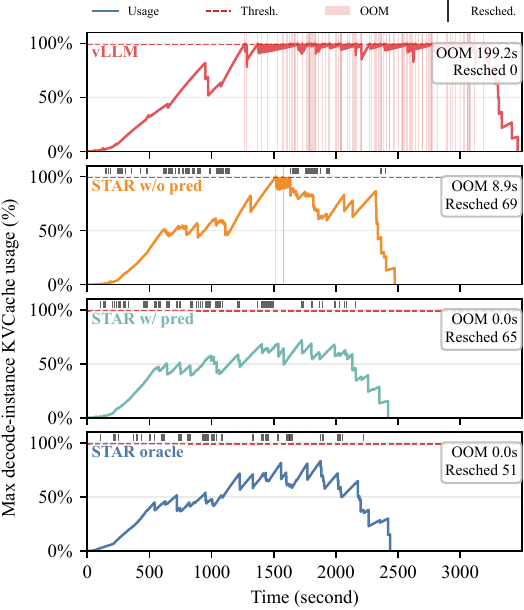}
    \caption{Runtime traces on the small cluster. The curve shows the maximum decode-instance KVCache usage; the dashed line marks the 99\% threshold; shaded regions indicate when OOM occurs; vertical ticks denote rescheduling events.}
    \label{fig:oom_reschedule_timeline}
    \vspace{-4pt}
\end{figure}

\subsection{Runtime Trace Analysis}

Figure~\ref{fig:oom_reschedule_timeline} presents runtime traces extracted from the same small-cluster experiment as the execution-on-small-scale study in Section~\ref{subsec:large_scale_simulation}. \textit{vLLM} remains near saturation for extended periods and repeatedly experiences OOM. \textit{\ours w/o prediction} substantially reduces OOM occurrences, whereas \textit{\ours w/ prediction} and \textit{\ours Oracle} stay below the 99\% threshold throughout the trace. These traces show that STAR improves goodput and P99 TPOT by better balancing decode load. Rescheduling suppresses severe imbalance, while prediction enables more forward-looking migrations and achieves a better balance with fewer unnecessary migrations.

\subsection{Prediction Robustness Analysis}

\begin{table}[htbp]
    \centering
    \caption{Prediction accuracy sensitivity on the large cluster.}
    \label{tab:prediction_quality_sensitivity}
    \begin{tabular}{@{}l|cccc@{}}
        \toprule
        Setting  & Exec. Var. & P99 TPOT & Goodput & Goodput Gain \\
        \midrule
        Full     & 0.163    & 26.49    & 0.157   & 10.56\%      \\
        6-bin    & 0.188    & 26.91    & 0.155   & 9.15\%       \\
        4-bin    & 0.220    & 27.70    & 0.148   & 4.23\%       \\
        2-bin    & 0.302    & 31.47    & 0.142   & 0.00\%       \\
        No pred. & 0.322    & 31.72    & 0.142   & 0.00\%       \\
        \bottomrule
    \end{tabular}
    \vspace{-1pt}
\end{table}

\begin{table}[htbp]
    \centering
    \caption{Prediction-interval tradeoff on the large cluster.}
    \label{tab:prediction_interval_tradeoff}
    \begin{tabular}{@{}l|cccc@{}}
        \toprule
        Interval & Exec. Var. & P99 TPOT & Goodput & Goodput Gain \\
        \midrule
        1 iter   & 0.237    & 27.84    & 0.148   & 4.23\%       \\
        20 iter  & 0.163    & 26.49    & 0.157   & 10.56\%      \\
        100 iter & 0.242    & 29.43    & 0.145   & 2.11\%       \\
        No pred. & 0.322    & 31.72    & 0.142   & 0.00\%       \\
        \bottomrule
    \end{tabular}
    \vspace{-1pt}
\end{table}

\stitle{Prediction Accuracy Sensitivity.}
We evaluate prediction accuracy sensitivity on the same large cluster setting as Figure~\ref{fig:overall_performance}. Specifically, the 2-bin, 4-bin, and 6-bin settings partition the remaining-length target into [0, 8K), [8K, 32K]; [0, 4K), [4K, 8K), [8K, 16K), [16K, 32K]; and [0, 2K), [2K, 4K), [4K, 6K), [6K, 8K), [8K, 16K), [16K, 32K], respectively. We use these non-uniform bins to align with the scheduler's main decision boundary: whether a request is near completion and unlikely to benefit from migration, or still long enough that migration is worthwhile. Here, \textit{Exec. Var.} denotes the execution-time variance across decode instances, consistent with Section~\ref{subsec:large_scale_simulation}. Table~\ref{tab:prediction_quality_sensitivity} shows that performance degrades gradually as the prediction granularity becomes coarser. The 6-bin predictor retains most of the benefit of the full predictor, with 0.155 vs. 0.157 goodput, only 0.42\,ms higher P99 TPOT, and only slightly higher execution-time variance (0.188 vs. 0.163). In contrast, the 2-bin predictor becomes nearly indistinguishable from \textit{No pred.}, with similarly high execution-time variance (0.302 vs. 0.322), indicating that STAR does not require exact token-count regression but does require sufficient granularity to separate substantially different remaining workloads.

\stitle{Prediction-Interval Tradeoff.}
Table~\ref{tab:prediction_interval_tradeoff} reports the effect of the reprediction interval under the same large cluster setting. A moderate interval of 20 decode iterations gives the best overall result. Repredicting every iteration increases prediction overhead and can trigger unnecessary migrations, while predicting every 100 iterations makes the scheduling decisions stale.

\section{Conclusion}
In this paper, we propose \ours, a decode-phase rescheduling framework, which integrates 1) a lightweight and accurate generation length predictor and 2) a prediction-based rescheduling algorithm to proactively mitigate workload imbalance in decode instances.
Evaluations under diverse workloads demonstrate that \ours significantly enhances load balance and achieves up to 2.63$\times$ higher goodput compared to existing systems. Furthermore, large-scale system simulations further validate the effectiveness of rescheduling and prediction for load balancing across clusters with up to 256 instances.

\begin{acks}
We sincerely thank the reviewers for their valuable comments and our shepherd Nikhil Jain for the helpful guidance. This work was supported by the National Natural Science Foundation of China under Grant No.~62572225, 62502198, 62325205, 62272215, and U25B2035; the Natural Science Foundation of Jiangsu Province under Grant No.~BK20251224; the Nanjing ``U35'' Talent Cultivation Program (No.~U (2024) 001); and the Fundamental and Interdisciplinary Disciplines Breakthrough Plan of the Ministry of Education of China (No.~JYB2025XDXM118). Rong Gu is the corresponding author of this paper.
\end{acks}

\bibliography{reference}

\end{document}